\documentclass[sn-nature,Numbered]{sn-jnl}

\usepackage{graphicx}%
\usepackage{multirow}%
\usepackage{amsmath,amssymb,amsfonts,bm}%
\usepackage{amsthm}%
\usepackage{mathrsfs}%
\usepackage[title]{appendix}%
\usepackage{xcolor}%
\usepackage{textcomp}%
\usepackage{manyfoot}%
\usepackage{booktabs}%
\usepackage{algorithm}%
\usepackage{algorithmicx}%
\usepackage{algpseudocode}%
\usepackage{listings}%
\usepackage{orcidlink}

\usepackage{subfig}
\usepackage{enumerate}
\usepackage[shortlabels]{enumitem}
\usepackage{mathtools}
\usepackage{bbold}
\usepackage[linesnumbered,ruled,vlined,algo2e]{algorithm2e}

\newcommand{\dprod}[2]{\langle #1,#2\rangle}
\newcommand{\dcos}{d_{\cos}}
\newcommand{\dspatial}{d_{\text{S}}}
\newcommand{\lambdas}{\lambda_{\text{S}}}
\newcommand{\lambdac}{\lambda_{\text{C}}}
\newcommand{\lambdaa}{\lambda_{\text{A}}}
\newcommand{\lambdag}{\lambda_{\text{G}}}

\newcommand{\djs}{d_{\text{JS}}}

\graphicspath{{figures/}}

\definecolor{ar_color}{RGB}{204,51,153}
\definecolor{jt_color}{RGB}{0,200,50}
\definecolor{js_color}{RGB}{0,134,139}

\newcommand{\smallspace}{\;\;}

\newcommand{\st}{\text{s.t.}\smallspace}

\def\emC{{C}}
\def\emD{{D}}

\def\emK{{K}}

\def\emM{{M}}

\def\emP{{P}}

\def\emT{{T}}

\def\emX{{X}}
\def\emY{{Y}}
\def\emZ{{Z}}

\def\1{\bm{1}}

\def\mC{{\bm{C}}}
\def\mD{{\bm{D}}}

\def\mI{{\bm{I}}}

\def\mM{{\bm{M}}}

\def\mP{{\bm{P}}}

\def\mT{{\bm{T}}}

\def\mX{{\bm{X}}}
\def\mY{{\bm{Y}}}
\def\mZ{{\bm{Z}}}

\def\mDelta{{\bm{\Delta}}}

\def\vzero{{\bm{0}}}

\def\vrho{{\bm{\rho}}}
\def\va{{\bm{a}}}
\def\vb{{\bm{b}}}

\def\ve{{\bm{e}}}

\def\vp{{\bm{p}}}
\def\vq{{\bm{q}}}
\def\vr{{\bm{r}}}

\def\vx{{\bm{x}}}

\def\sC{{\mathbb{C}}}

\def\sG{{\mathbb{G}}}

\def\sI{{\mathbb{I}}}

\def\sK{{\mathbb{K}}}

\def\sP{{\mathbb{P}}}

\def\sX{{\mathbb{X}}}

\def\gO{{\mathcal{O}}}

\newcommand{\KL}[2]{%
  D_{\text{KL}}\left(#1 \middle\| #2\right)%
}

\newcommand{\methodname}{DOT}
\newcommand{\methodnamet}{\texttt{\methodname}}

\newtheorem{proposition}{Proposition}
\newtheorem*{proposition*}{Proposition}

\newtheorem*{lemma*}{Lemma}

\newtheorem{remark}{Remark}

\DeclareMathOperator*{\argmin}{arg\,min}

\newcommand{\uS}{\text{S}}
\newcommand{\uR}{\text{R}}

\usepackage{setspace}
\begin{document}
\title{\methodname{}: A flexible multi-objective optimization framework for transferring features across single-cell and spatial omics}
\author[1,2]{\fnm{Arezou} \sur{Rahimi}\orcidlink{0000-0003-0285-8089}}\email{arezou.rahimi@uni-heidelberg.de}

\author[2]{\fnm{Luis} \sur{Vale Silva}\orcidlink{0000-0002-9658-5335}}\email{luis.a.valesilva@gsk.com}

\author[2]{\fnm{Maria} \sur{F\"{a}lth Savitski}\orcidlink{0000-0002-4923-3214}}\email{maria.x.faelth-savitski@gsk.com}

\author*[1,3]{\fnm{Jovan} \sur{Tanevski}\orcidlink{0000-0001-7177-1003}}\email{jovan.tanevski@uni-heidelberg.de}
\equalcont{These authors contributed equally to this work.}

\author*[1]{\fnm{Julio} \sur{Saez-Rodriguez}\orcidlink{0000-0002-8552-8976}}\email{pub.saez@uni-heidelberg.de}
\equalcont{These authors contributed equally to this work.}

\affil*[1]{\orgdiv{Institute for Computational Biomedicine}, \orgname{Heidelberg University \& Heidelberg University Hospital}, \orgaddress{Germany}}

\affil[2]{\orgdiv{Cellzome GmbH}, \orgname{GlaxoSmithKline}, \orgaddress{Heidelberg, Germany}}

\affil[3]{\orgdiv{Department of Knowledge Technologies}, \orgname{Jo\v{z}ef Stefan Institute}, \orgaddress{Ljubljana, Slovenia}}




\maketitle

\section*{Abstract}
Single-cell RNA sequencing (scRNA-seq) and spatially-resolved imaging/sequencing technologies have revolutionized biomedical research. On one hand, scRNA-seq provides information about a large portion of the transcriptome for individual cells, but lacks the spatial context. On the other hand, spatially-resolved measurements come with a trade-off between resolution and gene coverage. Combining scRNA-seq with different spatially-resolved technologies can thus provide a more complete map of tissues with enhanced cellular resolution and gene coverage. 
Here, we propose \methodname{}, a novel multi-objective optimization framework for transferring cellular features across these data modalities. \methodname{} is flexible and can be used to infer categorical (cell type or cell state) or continuous features (gene expression) in different types of spatial omics.
%
%
%
Our optimization model combines practical aspects related to tissue composition, technical effects, and integration of prior knowledge, thereby providing flexibility to combine scRNA-seq and both low- and high-resolution spatial data.
Our fast implementation based on the Frank-Wolfe algorithm achieves state-of-the-art or improved performance in localizing cell features in high- and low-resolution spatial data and estimating the expression of unmeasured genes in low-coverage spatial data across different tissues. \methodname{} is freely available and can be deployed efficiently without large computational resources; typical cases-studies can be run on  a laptop, facilitating its use.

\vtop{%
  \vskip-1ex
  \hbox{%
    \includegraphics[width=\textwidth]{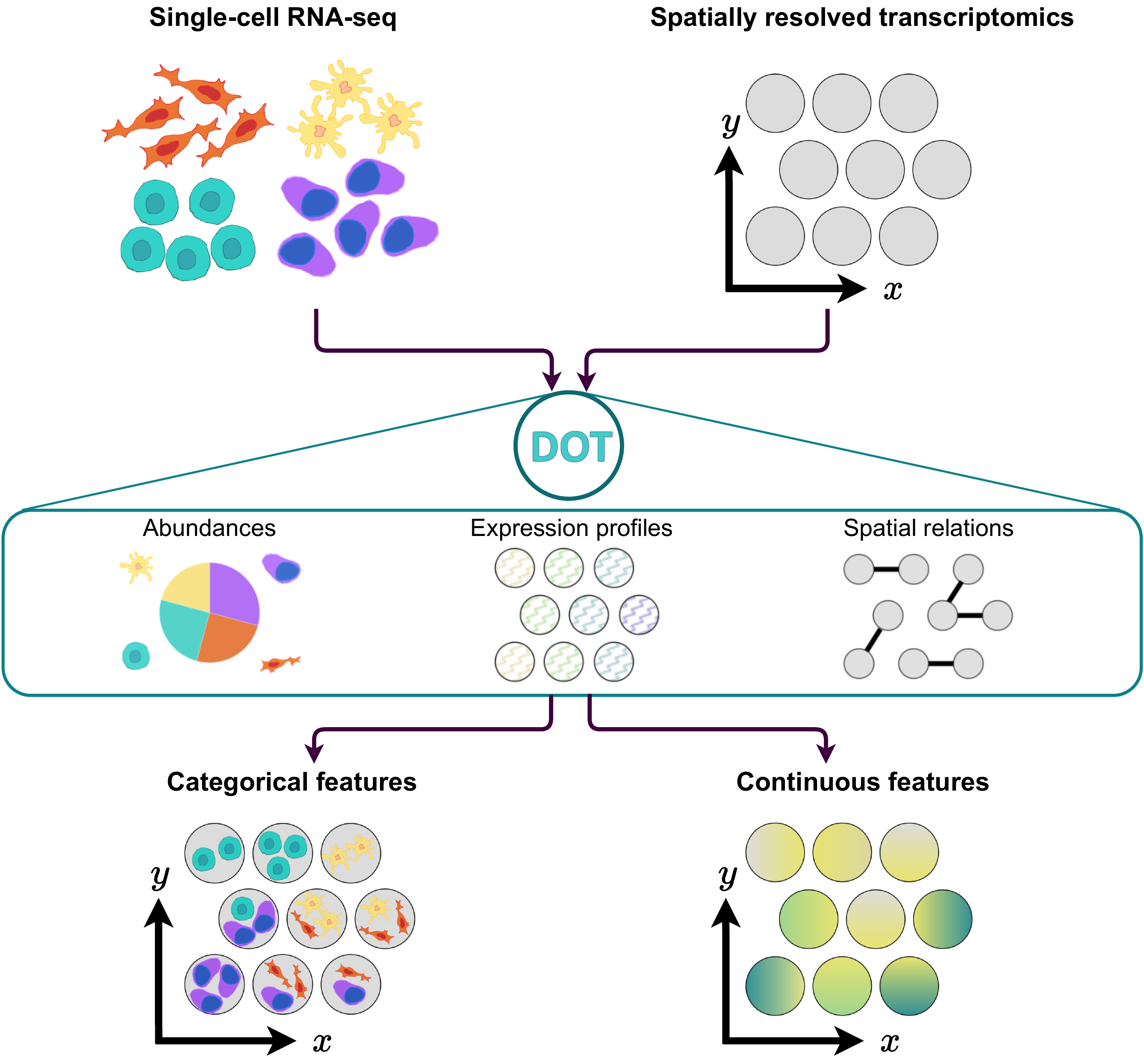}%
  }%
}%

\section{Main}

The organization of cells within human tissues, their molecular programs and their response to perturbations are central to better understand physiology, disease progression and to eventual identification of targets for therapeutic intervention \cite{trapnell2015defining,arendt2016origin}. 
%
%
Single-cell RNA sequencing can profile  a large part of the transcriptome of large portions of individual (single) cells. This has made these technologies (hereafter scRNA-seq) an essential tool for revealing distinct cell features (such as cell lineage and cell states) in complex tissues and has profoundly impacted our understanding of biological processes and the underlying mechanisms that control cellular functions \cite{papalexi2018single, cao2019single,rajewsky2020lifetime}. However, scRNA-seq requires dissociation of the tissue \cite{lee2020single}, losing the information about the spatial context and physical relationship between cells, that is critical to  understand the functioning of tissues.

To overcome these limitations, there has been recent advancements in spatially resolved transcriptomics (SRT) methods \cite{marx2021method, larsson2021spatially,rao2021exploring}. SRT methods measure gene expression in locations coupled with their two- or three-dimensional position. SRT methods vary in two axes: spatial resolution and gene coverage. On one hand, technologies such as Multiplexed Error-Robust Fluorescence In-Situ Hybridization (MERFISH) and In-Situ Sequencing (ISS), achieve cellular or even subcellular resolution \cite{chen2015spatially}, 
but are limited to measuring up to a couple of hundred pre-selected genes.
On the other hand, spatially resolved RNA sequencing, such as Spatial Transcriptomics  \cite{staahl2016visualization}, commercially available as 10X's Visium, and Slide-seq \cite{rodriques2019slide}, enable high-coverage gene profiling by capturing mRNAs \textit{in-situ} but come at the cost of measuring these averaged within spots that include multiple cells. Thus, there is a trade-off between resolution and richness (gene coverage) of SRT data.

A natural strategy to provide a complete picture is to combine scRNA-seq data with high-resolution SRT to transfer dissociated cells to spatial locations or  generally to combine scRNA-seq with low-resolution SRT is to estimate the composition of cell types in each spot. 
Alternatively, we can attempt to enrich the high-resolution SRT by predicting the expression of unmeasured genes. 
%
Integrating scRNA-seq and SRT is challenging for many reasons such as the limited number of genes shared across these modalities, differences in measurement sensitivities across technologies, and high computational cost for large-scale datasets.
%
Recent methods mostly rely on the genes that are captured both by scRNA-seq and SRT without using the remaining genes captured in each modality, do not use the \textit{spatial} relationships between cells in the spatial data, are limited to high or low resolution spatial data either in application or their underlying assumptions, and in many cases come with high computation cost for large instances \cite{zeng2022statistical}.
In Section~\ref{sec:related_work} we discuss the related work in more details.
%
%
Neglecting the spatial context is equivalent to assuming random placement of spots in the space, which is in contrast to the established structure-function relationship of tissues \cite{rao2021exploring}. Considering only a subset of genes limits the applicability of these methods to cases where the two data sets share several informative genes, which might not be the case when different technologies are used for profiling, or when few genes are measured in the spatial data (e.g., in MERFISH). 

In this article, we present \methodname{}, a versatile and scalable optimization framework, to integrate scRNA-seq and SRT for localizing the cell features via a multi-criteria mathematical program. 
Our model does not require the expression profiles to be mRNA counts and is applicable to both high- and low-resolution SRT, in the form of inferring membership probabilities for the former and relative or absolute abundance of cell types in the latter. 
We adapt a generalization of Optimal Transport with a tailored objective to leverage spatial information and to go beyond the use of only genes that are expressed in both modalities at the same time. 
Our optimization model is novel in considering several practical aspects in a unified framework, including (i) spatial relations between different cell features, (ii) differences in measurement sensitivity of different technologies, (iii) heterogeneity of cell sub-populations, (iv) compositional sparsity and size of spatial locations at different spatial resolutions, and (v) incorporation of prior knowledge about expected abundance of cell features \emph{in situ}.
We present a very fast implementation for our model based on the Frank-Wolfe algorithm thereby ensuring scalability and efficient solvability in large-scale datasets.
\methodname{} has a broader application beyond cell type decomposition, including transferring continuous features such as expression of genes that are missing in SRT but present in scRNA-seq data. 
\methodname{} is freely available to facilitate its application and further development.

\section{Results}\label{sec:results}
\subsection{\methodname{} is a versatile multi-objective optimization model for integrating spatial and single-cell omics}\label{sec:dot_overview}

\begin{figure*}[t!]
    \centering
    \includegraphics[width=\textwidth]{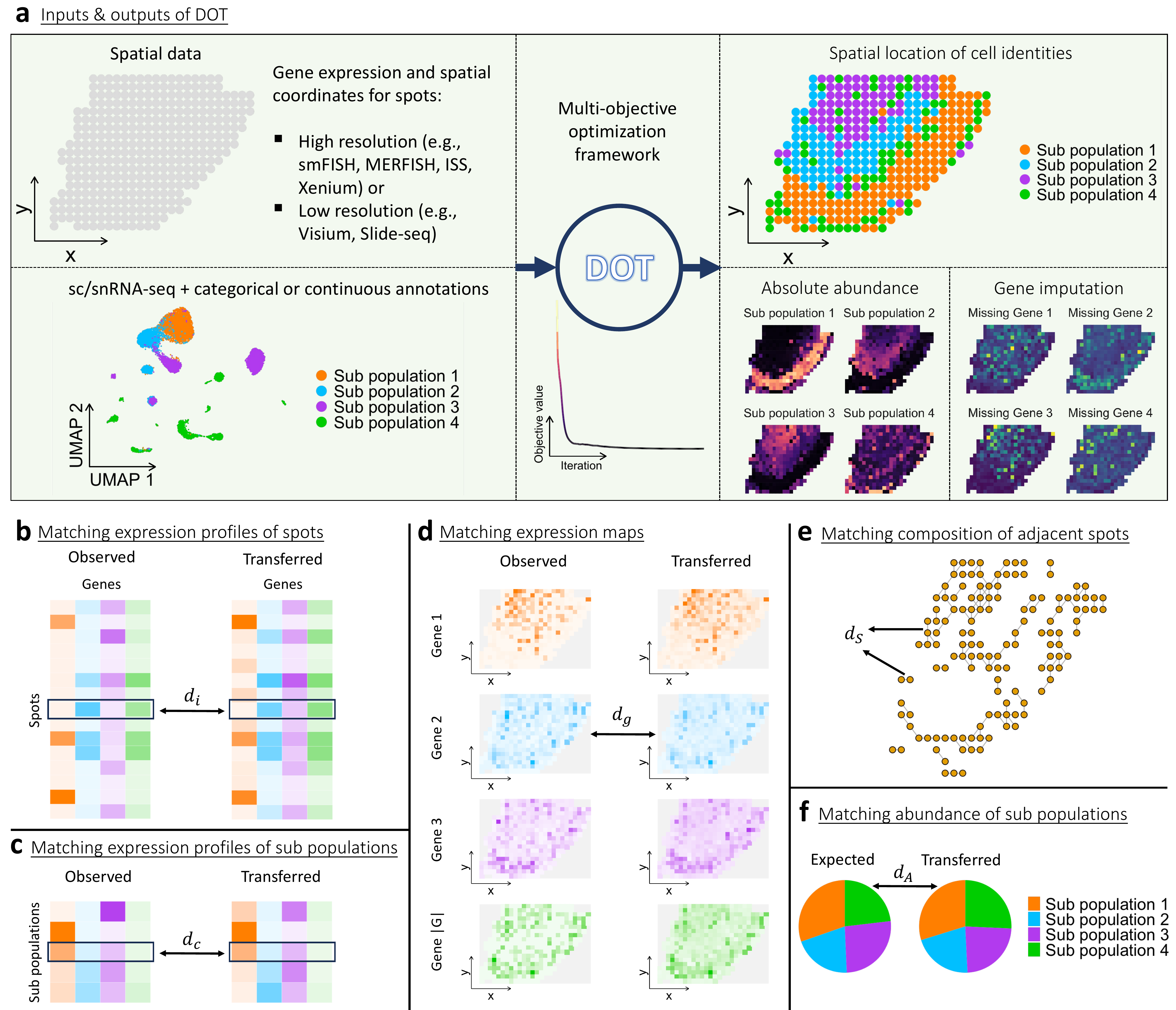}
    \caption{\textbf{Overview of inputs and outputs of \methodname{} and its optimization framework}. \textbf{a)} From left to right: \methodname{} takes two inputs: (i) spatially resolved transcriptomics data, which contains spatial measurements of genes at either high or low resolution spots and their spatial coordinates, and (ii) reference singe-cell RNA-seq data, which contains single cells with categorical (e.g., cell type) or continuous (e.g., expression of genes that are missing in the spatial data) annotations. \methodname{} employs several alignment objectives to locate the sub-populations and the annotations therein in the spatial data. The alignment objectives ensure a high quality transfer from different perspectives: 
    \textbf{b)} the expression profile of each spot in the spatial data (left) must be similar to the expression profile transferred to that spot from the reference data (right), 
    \textbf{c)} the expression profile of each sub population in the reference data (left) must be similar to the expression profile of that sub population inferred in the spatial data (right), 
    \textbf{d)} expression map of each gene in the spatial data (left) must be similar to expression map of that gene as transferred from the reference data (right), 
    \textbf{e)} spots that are both adjacent and have similar expression profiles should have similar compositions, 
    and \textbf{f)} if prior knowledge about the expected relative abundance of sub-populations is available, the transfer should retain the given abundances.}
    \label{fig:main}
\end{figure*}

Given a reference scRNA-seq data (R for short), which is a collection of single cells each annotated with a categorical or continuous feature (such as cell type), and a target spatially resolved transcriptomics data (S for short), which consists of a set $\sI$ of spots, associated with a location containing one or more cells, we wish to determine the abundances (in the  case of multiple cells per spot) or single value (in the case of a single cell per spot) of the unobserved feature(s) in spots of S (see Fig. \ref{fig:main}). 
%
In what follows, we assume that the unobserved features are categorical values in a set $\sC$ and note that the continuous case extends naturally. Consequently, we assume that the cells in R are categorized into $|\sC|$ sub-populations. 
Our mathematical model relies on determining a ``many-to-many'' mapping (transfer) $\mY$ of cell sub-populations in R to spots in S, with $\emY_{c,i}$ denoting the abundance of category $c\in \sC$ in spot $i\in \sI$. When S is high resolution, $\emY_{c,i}$ determines the probability that spot $i\in \sI$ is of type $c\in \sC$, whereas $\emY_{c,i}$ determines the absolute abundances when S is low resolution (i.e. spots are composed of multiple cells).

Let $\emX^{\uR}_{c,g}$ and $\emX^{\uS}_{i,g}$ denote the expression profiles of sub-population $c\in \sC$ and spot $i\in \sI$, respectively, for genes $g\in \sG$. We assume that $\emX^{\uR}_{c,g}$ is the mean expression of gene $g$ across the cells that belong to sub-population $c\in \sC$ of R (see Section~\ref{sec:multi-centroid} for extension to heterogeneous sub-populations). Moreover, $\emX^{\uS}_{i,g}$ is the aggregation of expression profiles of potentially several cells when S is low-resolution.
%
A high-quality transfer should naturally match the expression of the common genes across R and S. We ensure this by considering the following \textit{expression-focused} criteria:
\begin{enumerate}[(i)]
    \item \textit{Matching expression profile of spots} (Fig. \ref{fig:main}b). Expression profile of each spot $i\in \sI$ in S (i.e., $\mX^{\uS}_{i,:}$) should match the expression profile transferred to that spot from R via $\mY$ (i.e, $\sum\nolimits_{c\in \sC}\emY_{c,i}\mX^{\uR}_{c,:}$). We penalize the dissimilarity of these vectors via:
    \begin{align}
        d_{i}(\mY) \coloneqq \dcos(\mX^{\uS}_{i,:}, \sum\nolimits_{c\in \sC}\emY_{c,i}\mX^{\uR}_{c,:}). \label{eq:distance-location}
    \end{align}

    \item \textit{Matching expression profile of sub-populations} (Fig. \ref{fig:main}c). Expression profile of each sub-population $c\in \sC$ in R should match the expression profile of spots assigned to this sub-population via $\mY$:
    \begin{align}
        d_{c}(\mY) &\coloneqq \dcos(\mX^{\uR}_{c,:}, \sum\nolimits_{i\in \sI}\emY_{c,i}\mX^{\uS}_{i,:}). \label{eq:distance-centroid}
    \end{align}
    
    \item \textit{Matching gene expression maps} (Fig. \ref{fig:main}d). Expression map of each gene $g\in \sG$ in S should be similar to the expression map of that gene as transferred from R via $\mY$:
    \begin{align}
        d_{g}(\mY) \coloneqq \dcos (\mX^{\uS}_{:,g}, \sum\nolimits_{c\in \sC}\mY_{c,:}\emX^{\uR}_{c,g}). \label{eq:distance-gene}
    \end{align}
\end{enumerate}
In the above formulations, $\dcos$ is a scale-invariant metric based on cosine-similarity which measures the difference between two vectors regardless of their scales (Section~\ref{sec:distance-functions}).
In addition to the expression-focused objectives, we may incorporate prior knowledge in the form of the spatial location of spots as well as the expected abundance of cell sub-populations using the following \textit{compositional} criteria:
 \begin{enumerate}[resume*]
    \item \textit{Capturing spatial relations} (Fig. \ref{fig:main}e). Spots that occupy adjacent locations and have similar expression profiles are expected to be of similar compositions. Given $\sP$, the set of adjacent pairs of spots with similar expression profiles, we encourage similar composition profiles for these spots by penalizing
    \begin{align}
        \dspatial(\mY) \coloneqq \sum_{(i,j)\in \sP} w_{ij} \djs(\mY_{:,i}, \mY_{:,j}), \label{eq:distance-spatial}
    \end{align}
    where $\djs$ is the Jensen-Shannon divergence and $w_{ij}$ captures similarity of expression profiles of spots $i$ and $j$  (Section~\ref{sec:distance-functions}).
    \item \textit{Matching expected abundances} (Fig. \ref{fig:main}f). If prior information about the expected abundance of cell categories in S is available (e.g., when R and S correspond to adjacent tissues or consecutive sections), then abundance of cell categories transferred to S should be consistent with the given abundances. We measure dissimilarity between the vector of expected abundances (denoted $\vr$) and abundance of cell categories in S via
    \begin{align}
        d_{\text{A}}(\mY) \coloneqq \djs(\mY\ve, \vr).\label{eq:distance-abundance}
    \end{align}
\end{enumerate}

The expression-focused objectives naturally take precedence over the compositional objectives, especially when a large number of genes are common between R and S, but the compositional objectives are useful when the number of common genes is limited.
Note that objective (v) provides additional control over the abundance of cell types in S, but can be ignored if prior information about the abundance of cell types is not available. 

We treat these criteria as objectives in a multi-objective optimization problem and to consider them simultaneously (i.e., produce a Pareto-optimal solution), we optimize $\mY$ against a linear combination of these objectives as formulated below, hereafter referred as the \methodname{} model:
\begin{align}
    \min \quad & \sum_{i\in \sI} d_{i}(\mY)+\lambdac\sum_{c\in \sC} d_c(\mY)+\lambdag\sum_{g\in \sG} d_{g}(\mY)+\lambdas\dspatial(\mY)+
    \lambdaa d_{\text{A}}(\mY),  \label{dot:obj}
   \\
    \text{w.r.t.}\quad & \mY\in\mathbb{R}_+^{|\sC|\times |\sI|}, \label{dot:wrt}\\
    \text{s.t.}\quad & 1\le\sum\nolimits_{c\in \sC}\emY_{c,i}\le n_i \qquad \forall i\in \sI. \label{dot:cnt}
\end{align}
Here, $\lambdac$, $\lambdag$, $\lambdas$, and $\lambdaa$ are the user-defined penalty weights, and $n_i$ is an upper bound on the expected size (number of cells) of spot $i\in \sI$ (i.e., $n_i=1$ for high resolution SRT). For low-resolution SRT, we set $n_i=n$ for a pre-determined parameter $n$ and let the model determine the size of the spots (see Section~\ref{sec:setting}).


Next, we present an evaluation of the model, comparing its performance to the related work and highlight different aspects of \methodnamet{} in different applications. Briefly, we evaluate the performance of \methodnamet{} to transfer the cell type label of single-cell level spots in high-resolution SRT and decompose spots to cell type abundances in low-resolution SRT, and estimate the expression of genes that are missing in SRT but are measured in the reference scRNA-seq. Details of the datasets and performance metrics used for these experiments are presented in Appendix \ref{sec:data} and Section \ref{sec:metrics}, respectively.

\subsection{\methodname{} locates cell types in high-resolution spatial data}\label{sec:experiment-celltype-prediction}

Our goal with our first set of experiments is to evaluate the performance of different models in determining the abundance of cell types at each spot.
We used the high-resolution MERFISH spatial data of the primary motor cortex region (MOp) of the mouse brain \cite{zhang2021spatially}, which contains the spatial information and cell type of 280,186 cells across 75 samples (Appendix~\ref{sec:data_mop}).
Since the cell type represented in the spot is known in our high-resolution spatial data, we can use this information as ground truth when evaluating the performance of the different models. Details about the benchmark instances can be found in Section \ref{sec:data-preparing}.

\begin{figure}[t!]
    \centering
    \includegraphics[width=\textwidth]{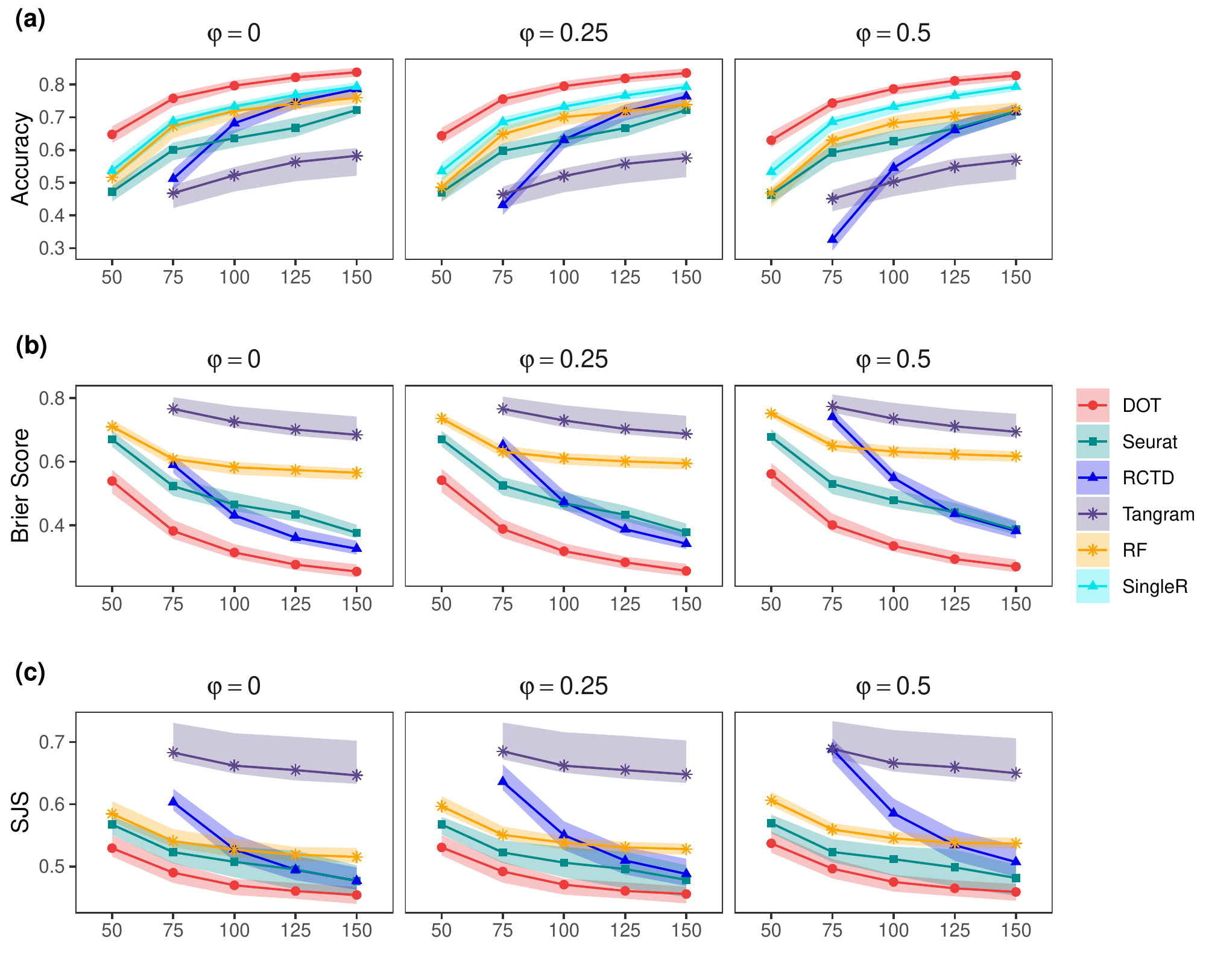}
    \caption{Performances of transfer of cell types in high-resolution spatial data as function of the gene coverage in the spatial data ($x$-axis) and as function of different amounts of noise in gene expression ($\varphi$). Points represent the median of 75 values, and the shaded areas correspond to their interquartile interval. 
    \texttt{SingleR} does not produce probabilities and is compared based on Accuracy only.
    }
    \label{fig:merfish_performances}
\end{figure}

We compared performance of \methodnamet{} against four models
from the literature: \texttt{RCTD} \cite{cable2021robust}, \texttt{Tangram} \cite{biancalani2021deep}, \texttt{Seurat} \cite{stuart2019comprehensive}, and \texttt{SingleR} \cite{aran2019reference} in transferring cell types from single-cell to high-resolution SRT. Given the multiclass classification nature of cell type prediction in high-resolution SRT, we also used \texttt{RF} \cite{breiman01ml} as a multiclass classifier baseline.

\methodnamet{} dominates the three specialized decomposition methods and the base line classification methods in assigning correct cell types to the spots (Fig.~\ref{fig:merfish_performances}a), and produces well-calibrated probabilities (Fig.~\ref{fig:merfish_performances}b) and better captures the relationships between cell types in space (Fig.~\ref{fig:merfish_performances}c), owing to its capacity to incorporate the spatial information through $\dspatial$. We also observe that even with very few genes in common between SRT and the reference scRNA-seq data (e.g., $|\sG|\le 75$), \methodnamet{} is able to reliably determine the cell type of spots in the space with high accuracy. In contrast, \texttt{RCTD} fails to produce results due to lack of shared information, and \texttt{Seurat} and \texttt{Tangram} produce results with low accuracy. The under-performance of \texttt{Seurat} is due to its over-fitting to the a prior distribution of cell types in the reference data, while \texttt{Tangram} struggles with the large number of cells in the reference data not being matched with the target spatial data. We also observe that \methodnamet{} performs robustly under fluctuations in the gene expression.

\subsection{\methodname{} determines cell type abundances in low-resolution spatial data}\label{sec:experiment-celltype-decomposition}

\begin{figure}[t!]
    \centering
    \begin{subfloat}[]{
         \includegraphics[width=0.44\textwidth, clip]{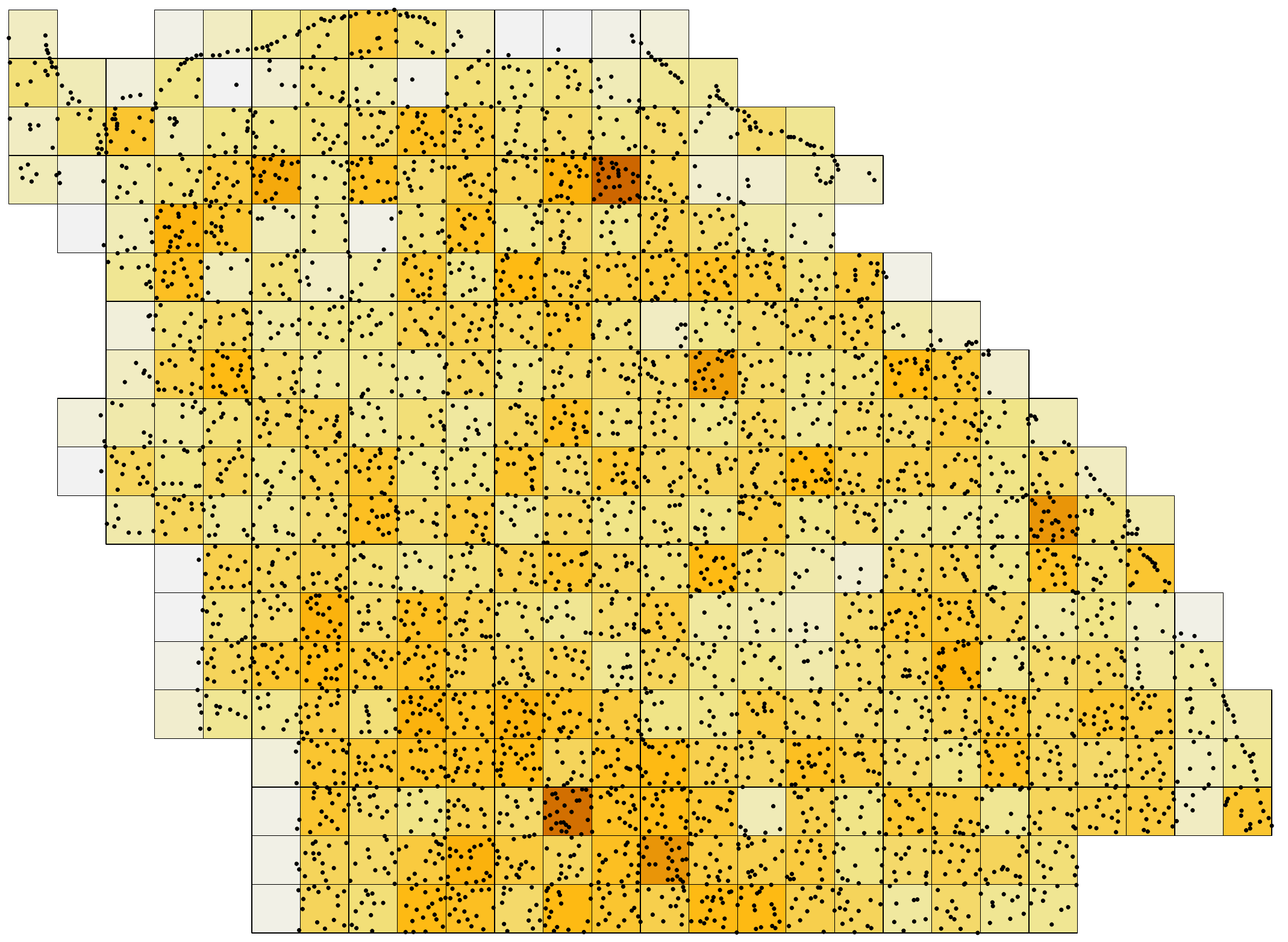}}
     \end{subfloat}
     \begin{subfloat}[]{
         \includegraphics[width=0.44\textwidth, clip]{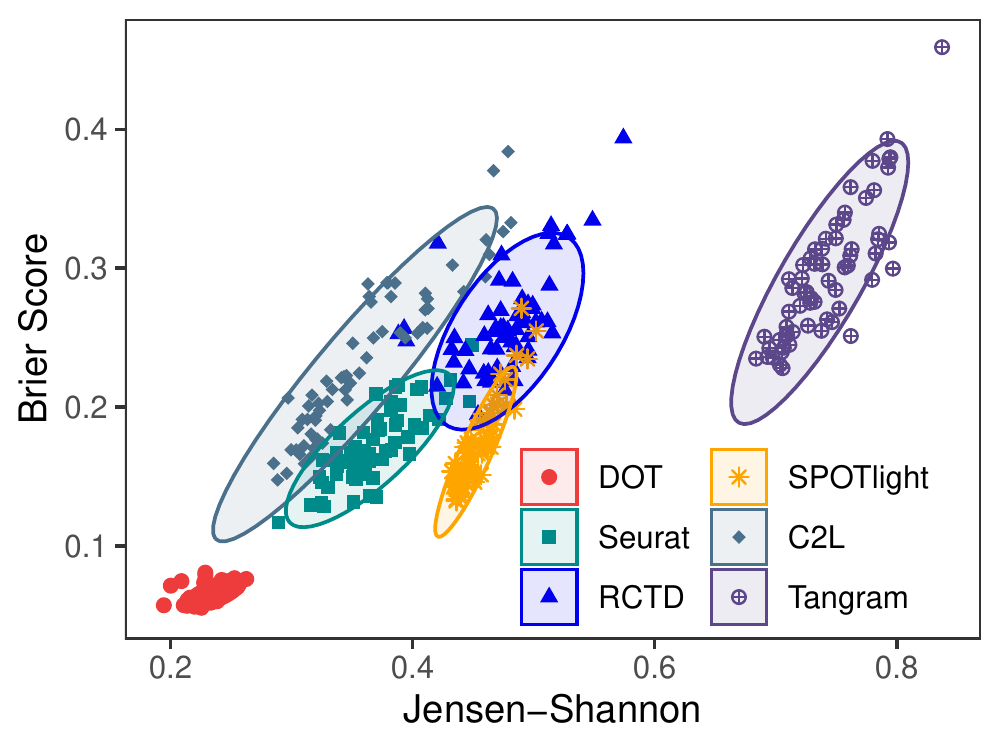}}
     \end{subfloat}
      \begin{subfloat}[]{
         \includegraphics[width=\textwidth, clip]{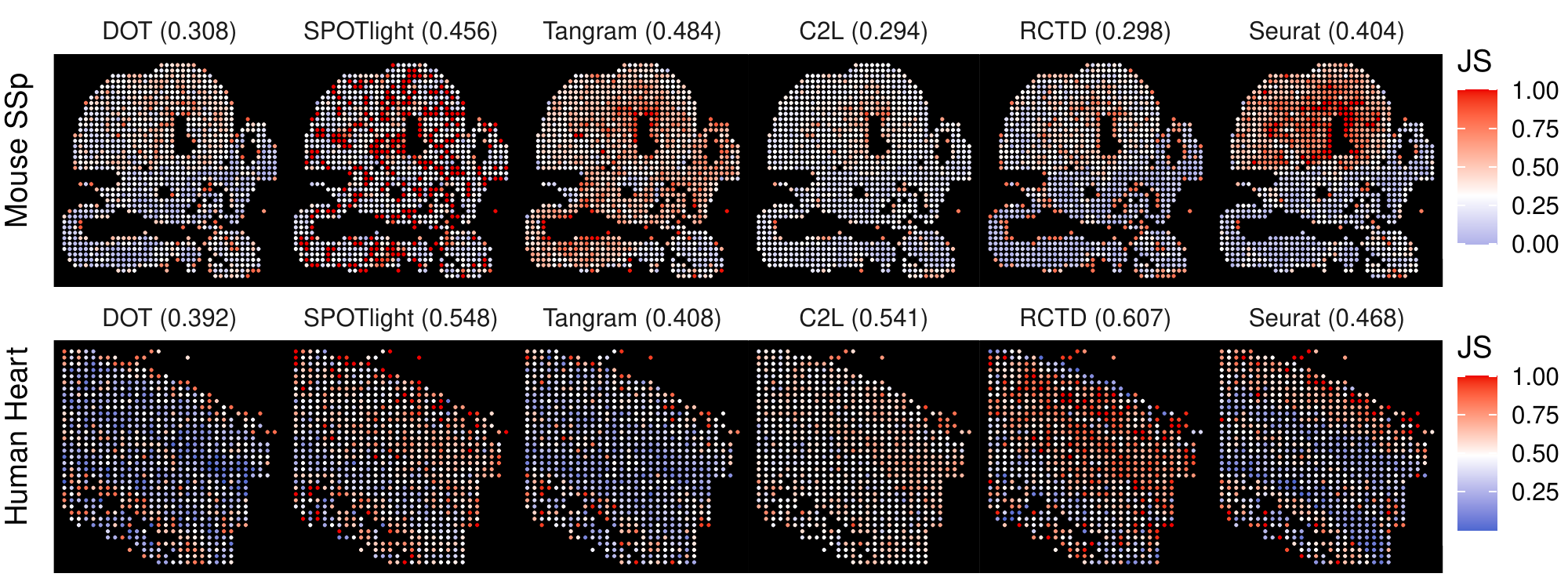}}
     \end{subfloat}
    \caption{(a) Synthetic low-resolution SRT from high-resolution SRT. Dots represent cells and tiles represent multicell spots. (b) Performance of the algorithms in the low-resolution spatial data across 75 samples of MOp. Each point denotes the average performance across all spots in the sample. (c) Distribution of performance of models on each individual spot in the low-resolution spatial data of Mouse SSp (top) and developing human heart (bottom). Each subplot shows the distribution of prediction error based on the Jensen-Shannon divergence at each spot in the spatial data, with the average value over all spots given on top of each plot. }
    \label{fig:multicell}
\end{figure}

Since there is no ground truth for real low-resolution spatial data such as Visium and Slide-seq, we produce ground truth low-resolution spatial data in an objective manner by reproducing measurements of low-resolution data by pooling adjacent cells in the high-resolution spatial data of primary motor cortex of the mouse brain (MOp), primary somatosensory cortex of the mouse brain (SSp), and the developing human heart. Fig.~\ref{fig:multicell}a illustrates a sample low-resolution SRT obtained from the high-resolution MERFISH data of a MOp tissue.

In Fig.~\ref{fig:multicell}b we show the comparison of the performance of \methodnamet{} against \texttt{RCTD}, 
\texttt{SPOTlight} \cite{elosua2021spotlight},
cell2location (\texttt{C2L}) \cite{kleshchevnikov2022cell2location}, \texttt{Tangram} and \texttt{Seurat} in determining the cell type composition of the multicell spots created based on the MOp dataset (see Section \ref{sec:data-preparing} for details on the benchmark instances). We observe that \methodnamet{} outperforms other models with respect to both Jensesn-Shannon and Brier Score metrics.

We next used single-cell level spatial data coming from osmFISH technology 
\cite{codeluppi2018spatial} to produce multicell data for SSp (Section \ref{sec:data_ssp}).
Subsequently, for the developing human heart, we used subcellular spatial data generated by the ISS platform \cite{asp2019spatiotemporal} (Section \ref{sec:data_heart}). 
We tested the performance of \methodnamet{} against the five deconvolution methods on these two samples, results of which are illustrated in Fig.~\ref{fig:multicell}c. \methodnamet{} outperforms other models in the human heart sample and is among the best-performing models in the mouse SSp sample. We also observe that \methodnamet{} exhibits a uniform performance across different regions of the tissues, which implies that the performance of \methodnamet{} is not sensitive to different regions/cell types of the tissue (compare to \texttt{Tangram} and \texttt{Seurat} in SSp and \texttt{RCTD} in human heart). These results further highlight the competitive performance of \methodnamet{} and its robustness in identifying the cell type composition of spots across different tissues.

\subsection{\methodname{} estimates the expression of unmeasured genes in spatially resolved data accurately}\label{sec:experiment-expression-estimation}

Given that in high-resolution SRT typically only a few genes are measured, the expression of genes that were not measured in SRT can be estimated by transferring scRNA-seq to SRT. Therefore,  we evaluate the performance of \methodname{} in estimating the expression of missing genes in the high-resolution SRT using the spatial data from breast cancer tumor microenvironment \cite{janesick2022high} (see Appendix~\ref{sec:data_breast_cancer}). 
As the high- and low-resolution SRT in this dataset come from the same tissue section, we can use the gene expression maps in low-resolution SRT as a proxy for ground truth to evaluate the expression maps of the missing genes in the high-resolution SRT as estimated by \methodname{}.

\begin{figure*}[t!]
     \centering
     \begin{subfloat}[]{
         \includegraphics[width=\textwidth, clip]{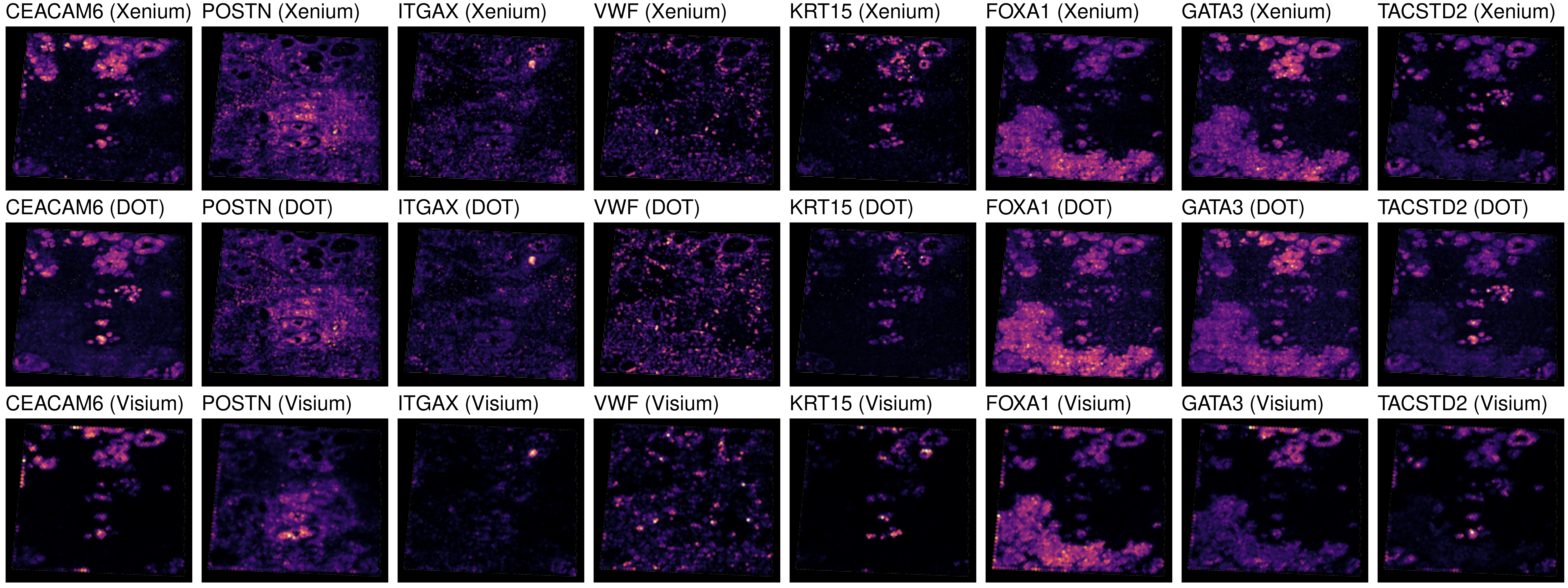}}
     \end{subfloat}
     \begin{subfloat}[]{
         \includegraphics[width=0.61\textwidth, clip]{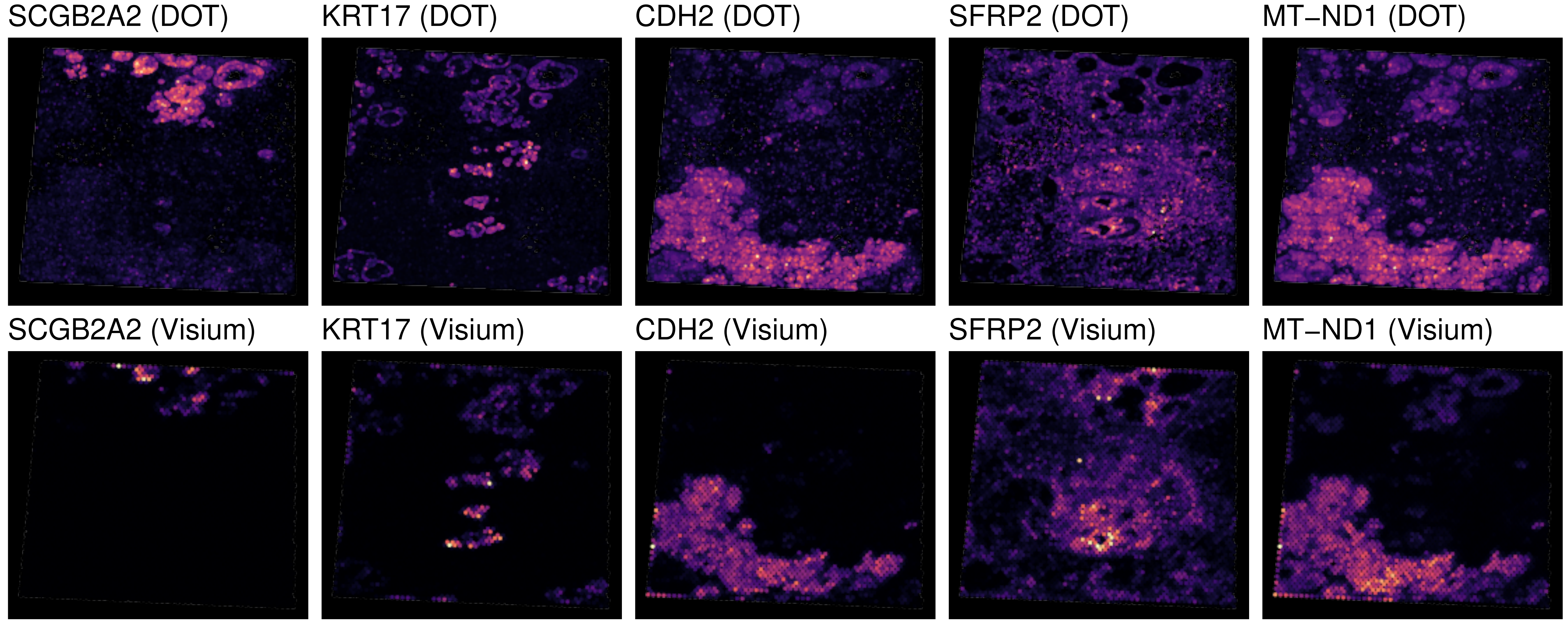}}
     \end{subfloat}
     \hspace{0.02\textwidth}
     \begin{subfloat}[]{
         \includegraphics[width=0.32\textwidth, clip]{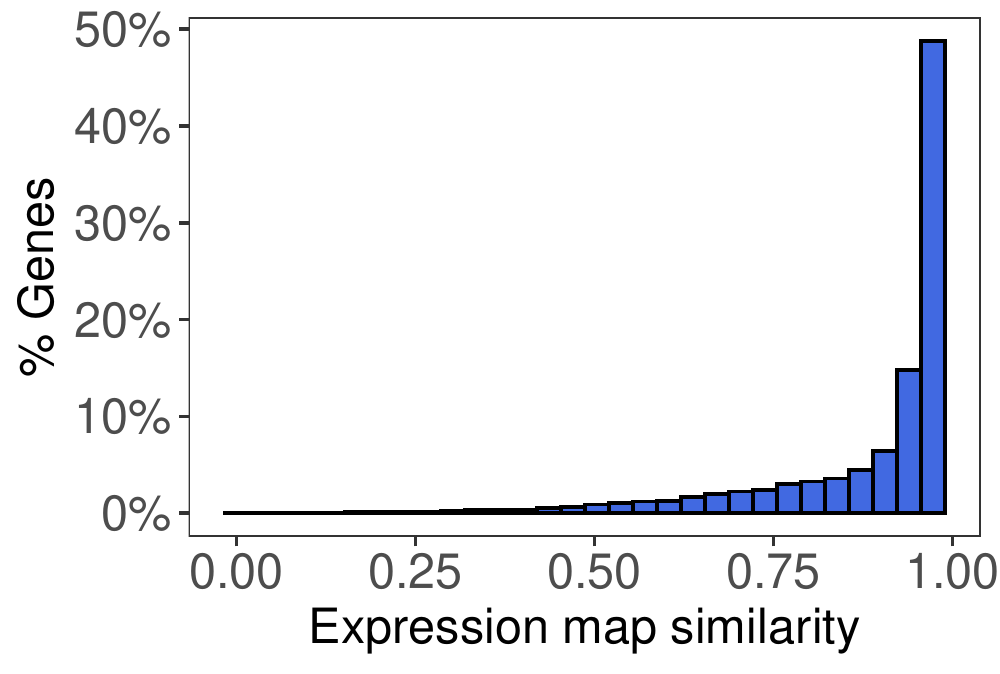}}
     \end{subfloat}
     \caption{(a) Expression map of eight breast cancer markers measured in both Xenium (ground truth; top) and Visium (low-resolution proxy; bottom), and as transferred from scRNA-seq to Xenium using \methodname{} (estimated; middle). Brighter means higher expression. (b) Expression map of five breast cancer markers that are measured in Visium (bottom) but are missing in Xenium and are transferred from scRNA-seq using \methodname{} (top). (c) Cosine similarity between expression maps of Visium and \methodname{} for the genes that are not measured in Xenium.}\label{fig:expression_map_xen_dot_vis}
\end{figure*}

We started by evaluating the performance of \methodname{} on genes that are present in the high-resolution spatial data as ground truth.  
In Fig.~\ref{fig:expression_map_xen_dot_vis}a we show a qualitative comparison of maps of eight genes related to breast cancer \cite{risom2022transition} produced by \methodname{} with those of high-resolution (ground truth) and low-resolution data (approximate ground truth). The expression maps produced by \methodname{} match almost perfectly with the ground truth expression maps. Both \methodname{} and the ground truth high-resolution spatial data also match the low-resolution gene expression maps almost perfectly, which further validate the quality of the solution produced by \methodname{}. Note that due to the single-cell resolution of the high-resolution spatial data colors are brighter. Nonetheless, the spatial patterns match between all three rows.  

Fig.~\ref{fig:expression_map_xen_dot_vis}b illustrates the expression maps of five genes associated with breast cancer that are not measured in the high-resolution spatial data but are estimated by \methodname{}. For a quantitative comparison of expression maps in the high- and low-resolution SRT, given that there is no one-to-one correspondence between single-cell spots in the high-resolution and multicell spots in the low-resolution spatial data, we split the tissue into a 10 by 10 grid, and aggregated the expression of each gene within each tile.
Consequently, we obtained two 100 by 18,000 matrices, one for the ground truth low-resolution spatial data and another for \methodname{}. Fig.~\ref{fig:expression_map_xen_dot_vis}c compares the column-wise cosine similarities across different genes.
These results further confirm the ability of \methodname{} in reliably estimating the expression of missing genes in high-resolution spatial data.

\subsection{\methodname{} is efficient and scalable}

We designed the mathematical model and the solution method for \methodname{} with particular attention to scalability and computational efficiency. 
In terms of algorithmic performance (Table~\ref{tab:all_time}), \methodnamet{} takes on average 426 seconds to solve each instance of the high resolution spatial data, which is an order of magnitude faster than \texttt{RCTD}, \texttt{Tangram}, and \texttt{RF}, and is comparable to \texttt{Seurat} and \texttt{SingleR}.
Similarly, \methodnamet{} took on average 433 seconds to solve the low-resolution instances of MOp, which proved to be more than twice faster than \texttt{Seurat}, and orders of magnitude faster than \texttt{RCTD}, \texttt{SPOTlight}, \texttt{C2L} and \texttt{Tangram}, further highlighting the superiority of \methodnamet{} in terms of both accuracy and computational efficiency.

\begin{table*}[h]
\centering
\resizebox{\textwidth}{!}{%
\begin{tabular}{@{}ccrrrrrrrrr@{}}
\toprule
Experiment & Resolution & Instances
& \textbf{\methodname} & \textbf{Seurat}  & \textbf{RCTD} & \textbf{Tangram} & \textbf{SPOTlight} &
\textbf{C2L} &\textbf{SingleR} & \textbf{RF} \\ \midrule
MOp & High & 1125 &
426 & 380 & 4748 & 10141 & 7884 & 3310 & 303 & 7427\\ 
MOp & Low & 75 &
433 & 1086 & 4705 & 8250 & 52825 & 6119 & --- & --- \\
SSp & Low & 1 &
4 & 21 & 117 & 248 & 705 & 364 & --- & --- \\
Heart & Low & 1 &
8 & 11 & 185 & 88 & 316 & 398 & --- & --- \\
\bottomrule
\end{tabular}%
}
\caption{Average computation times (in seconds) across different experiments.}
\label{tab:all_time}
\end{table*}

\section{Discussion}
Single-cell RNA-seq and spatially-resolved imaging/sequencing technologies provide each a partial picture in understanding the organization of complex tissues. To obtain a full picture, computational methods are needed to  combine  these two data modalities. 

We present \methodname{}, a versatile, fast and scalable optimization framework for transferring cell sub-populations from a reference scRNA-seq data to tissue locations, thereby transferring categorical and continuous features from the reference data to the spatial data. \methodname{} can help to improve our understanding of cellular functions and tissue architecture. Our optimization framework employs several alignment measures to assess the quality of transfer from different perspectives and determines the relative or absolute abundance of different sub-populations \emph{in situ} by combining these metrics in a multi-objective optimization model. 
Our metrics are designed to account for potentially different gene expression scales across the two modalities.
Moreover, based on the premise that nearby locations with similar expression profiles posses similar compositions, our model leverages the spatial information as well as both joint and dataset-specific genes in addition to matching the expression of common genes. In addition, whenever prior information about the abundance of cell features in the spatial data is available (e.g., estimated from a similar tissue), our model gives the user the flexibility to match these abundances to a desired level. Our model also takes into account inherent heterogeneity of cell sub-populations through a pre-processing step to ensure that refined sub-clusters of the reference are transferred.

Our model is applicable to both high-resolution (such as MERFISH) and low-resolution (such as Visium) spatial data and can be used for gene intensity or expression count data. While we use the same optimization framework for both high- and low-resolution spatial data, our model has specific  features to account for the distinct features of these modalities. In particular, our model can determine the size of spots in low-resolution spatial data and accounts for sparsity of composition of spots. For instance, in the context of inferring cell type composition of spots, this allows us  to produce pure cell type compositions for high-resolution spatial data and mixed compositions for low-resolution spatial data.

While our optimization model in its most general form involves several components, we have designed a solution method based on the Frank-Wolfe algorithm with special attention to scalability to large-scale reference and spatial data. Moreover, our implementation reduces involvement of the user in parameter tuning by estimating the objective weights and other hyper parameters of the model from the data, thereby facilitating application of \methodname{} to different problems with minimal implementation effort. Given that our model theoretically generalizes optimal transport (see Section \ref{sec:related_work} and Appendix~\ref{app:connection-to-ot}), we envision that \methodname{} can be integrated with OT-based computational frameworks such as \textit{moscot} \cite{klein2023mapping} in the future.

Using experiments on data from mouse brain, human heart, and breast cancer, we showed that \methodname{} predicts the cell type composition of spots and expression of genes in spatial data with high accuracy, achieving and often outperforming the state-of-the-art methods both in terms of predictive performance and computation time. Although we demonstrated the application of \methodname{} in transferring cell type labels and inferring the expression of missing genes, our model can be used for transferring other features such as Transcription Factor and pathway activities inferred from the reference scRNA-seq data \cite{holland2020robustness}. Additionally, our optimization framework can potentially be extended to alignment of spatial multiomics by exploiting the spatial information of the different data types. As our formulation is hypothesis-free (i.e., does not rely on statistical assumptions based on mRNA counts), \methodname{} naturally extends to applications in other omics technologies.

\section{Methods}

\subsection{Related work}\label{sec:related_work}
%
Several decomposition methods (also known as deconvolution methods) have been proposed in recent years \cite{zeng2022statistical}. 
As cell type decomposition, particularly in the high-resolution spatial data, is inherently a multiclass classification task, classification methods, such as Random Forests \cite{breiman01ml}, can be used for tackling this problem. However, because of the domain-specific properties of this problem, including differences in gene coverage, resolution, measurement sensitivity, and modality-specific characteristics, there has been an increased interest in improvement and new method development to aggregate scRNA-seq and SRT since the initial efforts \cite{tanevski2020,palla2022spatial}.

%
SPOTlight \cite{elosua2021spotlight} uses non-negative matrix factorization regression to factorize the scRNA-seq count matrix into \textit{topic profile} and \textit{topic distribution} matrices. SPOTlight then uses non-negative least squares regression to model the gene expressions in spots as a product of the topic profile matrix learned from scRNA-seq and a topic distribution matrix, which is then used to determine the cell type composition of spots.
Robust cell type decomposition (RCTD) \cite{cable2021robust} fits a statistical model by maximum-likelihood estimation, assuming a Poisson distribution for the expression of each gene at each spot. 
%
Cell2location is another statistical model which assumes a two-step Bayesian model for inferring cell type composition of spots  \cite{kleshchevnikov2022cell2location}. 
In the first step, it estimates reference cell type centroids from single-cell profiles. In the second step, cell2location uses these reference centroids to decompose mRNA counts at individual spatial locations into reference cell types.

While the aforementioned methods are designed specifically for low-resolution spatial data, some are also applicable to high-resolution spatial data.
Among the methods that are specialized for high-resolution spatial data, 
Tangram \cite{biancalani2021deep} incorporates a deep learning model to find the best placement of single cells in spots using a designed loss function and can thus carry cell type information as a byproduct. 
Seurat V3 workflow  \cite{stuart2019comprehensive} is a widely-used toolkit for analyzing scRNA-seq data, which offers an ``anchoring'' technique based on mutual nearest neighbours classifier for aligning two modalities in the space of principal components.

From a methodological standpoint, our formulation generalizes Optimal Transport (OT) (see Appendix~\ref{app:connection-to-ot}), which is a way to match, with minimal cost, data points between two domains embedded in possibly different spaces using different variants of the Wasserstein distance \cite{villani2021topics,santambrogio2015optimal, peyre2019computational,zhang2019optimal}.
Over the past years, OT has been applied to various machine learning problems in a wide variety of contexts such as
generative modeling \cite{bunne2019learning}, 
feature aggregation
\cite{mialon2020trainable}, 
dataset denoising \cite{wang2022optimal},
generalization error prediction
\cite{chuang2021measuring}, 
graph matching/classification \cite{titouan2019optimal}, 
and domain adaptation \cite{li2021divergence}. In particular, OT has been employed in computational biology with applications such as transporting entities from one cross sectional measurement to the next using unbalanced dynamic transport \cite{tong2020trajectorynet}, studying developmental time courses and understanding the molecular programs that guide differentiation during development \cite{schiebinger2019optimal}, reconstructing developmental trajectories from time courses with snapshots of cell states and lineages \cite{forrow2021lineageot}, reconstructing the organization of cells in the tissue \cite{nitzan2019gene, cang2020inferring} and alignment of spatial omics \cite{zeira2022alignment}. In addition, computational pipelines with OT components have been developed to facilitate applications of OT in computational biology \cite{klein2023mapping}.

In Appendix~\ref{app:connection-to-ot} we establish the connections between our formulation and OT formulations, and highlight the distinct features of our model that make it more suitable for the task of transferring annotations from the reference sub-populations to high- or low-resolution spatial data. Briefly, we note that our distance functions $d_i$ and $\dspatial$ share elements with Fused Gromov-Wasserstein (FGW) \cite{vayer2020fused}, which is also implemented as part of \textit{moscot}  \cite{klein2023mapping}. Indeed, we present metrics for R and S for which the resulting FGW encourages similar compositions for adjacent spots with similar expression profiles, thereby its connection to our definition of set $\sP$ and our distance function $\dspatial$. 

Besides the specialized distance functions included in the objective function of \methodname{} that measure quality of the transport map from different practical perspectives, there are other substantial differences between the common components of our formulation and FGW. The first difference is that OT formulations, including FGW, construct their transportation cost matrix by assuming that each spot is assigned to exactly one sub-population, discarding the fact that spots in low-resolution spatial data are composed of multiple cells coming from potentially different sub-populations. In contrast, our $d_i$ distance captures both mixed and pure compositions. Moreover, scale invariance of $d_i$, together with our $d_c$ and $d_g$ distance functions, allow us to determine the size of spots as part of the optimization process, whereas OT variants require the sizes as given. It is also important to note that our spatial distance function $\dspatial$ is convex, and by design, scales in order $\gO(|\sI|\; |\sC|)$ (i.e., linearly in the number of spots and sub-populations), while FGW formulations are non-convex and scale in $\gO(|\sI|^2|\sC|+|\sC|^2|\sI|)$ \cite{vayer2020fused}, making \methodname{} more appealing from a computational view for large-scale datasets.

\subsection{Mathematical model}\label{sec:math_model}
\subsubsection{Deriving the distance functions}\label{sec:distance-functions}
To assess dissimilarity between expression vectors $\va$ and $\vb$, we introduce the distance function
\begin{align}
    \dcos(\va,\vb) \coloneqq \sqrt{1-\cos\left(\va,\vb\right)}, \label{eq:dcosine-metric}
\end{align}
where $\cos\left(\va,\vb\right) = \frac{1}{\|\va\|\|b\|}\dprod{\va}{\vb}$.
We note that, unlike cosine dissimilarity (i.e., $1-\cos(\cdot,\cdot)$), $\dcos$ is a \textit{metric} distance function.
Moreover, $\dcos$ is quasi-convex for positive vectors $\va$ and $\vb$, and is scale-invariant, in the sense that it is indifferent to the magnitudes of the vectors. This is by design, since we want to assess dissimilarity between expression vectors regardless of the measurement sensitivities of different technologies. When assessing the gene expression profiles, this also allows to measure the differences regardless of the size of spots and cell sub-populations. 

With this distance metric, by minimizing $d_i(\mY)$ as defined in Eq. \eqref{eq:distance-location}, we ensure that the vector of gene expressions in spot $i\in \sI$ (i.e., $\mX^{\uS}_{i,:}$) is most similar to the vector of gene expressions transferred to spot $i$ through $\mY$ (i.e., $\sum_{c\in \sC}\emY_{c,i}\mX^{\uR}_{c,:}$). Similarly, 
with $d_c(\mY)$ as defined in Eq. \eqref{eq:distance-centroid}, we minimize dissimilarity between centroid of sub-population $c\in \sC$ in R (i.e., $\mX^{\uR}_{c,:}$) and its centroid in S as determined via $\mY$, i.e., $\frac{1}{\rho_c}\sum_{i\in \sI}\emY_{c,i}\mX^{\uS}_{i,:}$, where $\rho_c=\sum_{i\in \sI}Y_{c,i}$ is the total number of spots in S assigned to $c$. Given the scale-invariance property of $\dcos$, we can drop $1/\rho_c$ and derive  Eq. \eqref{eq:distance-centroid} as
\begin{align*}
    d_{c}(\mY) &\coloneqq \dcos\left(\mX^{\uR}_{c,:}, \frac{1}{\rho_c}\sum\nolimits_{i\in \sI}\emY_{c,i}\mX^{\uS}_{i,:}\right)=\dcos\left(\mX^{\uR}_{c,:}, \sum\nolimits_{i\in \sI}\emY_{c,i}\mX^{\uS}_{i,:}\right).
\end{align*}
We also note that $d_g(\mY)$ as defined in Eq. \eqref{eq:distance-gene} measures the difference between the expression map of gene $g\in \sG$ in S (i.e., $\mX^{\uS}_{:, g}$) and the one transferred to S through $\mY$ (i.e., $\sum_{c\in \sC}\mY_{c,:}\emX^{\uR}_{c,g}$) regardless of the scale of the expression of $g$ in S and R up to a constant multiplicative factor.

Our goal with objective (iv) as defined in Eq. \eqref{eq:distance-spatial} is to leverage the spatial information and potentially features that are contained in S but not in R to encourage spots that are adjacent in the tissue and exhibit similar expression profiles to attain similar cell type compositions. (Note that we do not assume that all adjacent spots should attain similar cell type compositions.) To achieve this goal, we define $\sP$ as 
\begin{align}
    \sP = \left\{(i,j)\in \sI^2: w_{i,j}\ge \bar{w},\quad \|\vx_i-\vx_j\|\le \bar{d},\quad i < j\right\} \label{eq:spatial-pairs}
\end{align}
to denote the set of pairs of spots $(i,j)$ that are adjacent ($\|\vx_i-\vx_j\|\le \bar{d}$) and have similar expression profiles ($w_{i,j}\ge \bar{w}$), with $\vx_i$ denoting the spatial coordinates of spot $i$ in  $\mathbb{R}^2$ or $\mathbb{R}^3$, and $w_{ij}=\cos(\mX^{\uS}_{i, :}, \mX^{\uS}_{j, :})$ denoting the cosine similarity of spots $i$ and $j$ according to the full set of genes measured in S (i.e., $\sG^{\uS}$). Here, $\bar{d}$ is a given distance threshold and $\bar{w}$ is a cutoff value for cosine similarity. As a larger $\bar{w}$ results in a smaller set $\sP$, we can ensure that $\dspatial$ can be computed linearly in the number of spots $|\sI|$ by choosing a proper value for $\bar{w}$ such that $|\sP|=\gO(|\sI|)$ (see also Section~\ref{sec:setting}).


We employ Jensen-Shannon divergence defined as 
\begin{align}
    \djs(\vp,\vq) = \frac{1}{2}\KL{\vp}{ \frac{\vp+\vq}{2}}+\frac{1}{2}\KL{\vq}{\frac{\vp+\vq}{2}}, \label{eq:jensen-shannon}
\end{align}
to measure dissimilarity between distributions $\vq$ and $\vp$,
where $\KL{\vp}{\vq} = \sum_{j}p_j\log(p_j/q_j)$ is the Kullback–Leibler divergence \cite{manning1999foundations}. We remark that $\djs(\vp,\vq)$ is strongly convex and does not require absolute continuity on distributions $\vq$ and $\vp$ \cite{gallager1968information}.

Finally, if prior information about the expected abundance of cell types in S is available (e.g., estimated from a neighboring single-cell level tissue), we denote the expected abundance of cell type $c\in \sC$ in S by $r_c$. Note that abundance of cell type $c\in \sC$ in S according to $\mY$ is $\rho_c \coloneqq \sum_{i\in \sI}\emY_{c,i}$. Since $\vr$ and $\vrho$ need not be mutually continuous, we employ $\djs(\vrho,\vr)$ in Eq. \eqref{eq:distance-abundance} to measure the difference between $\vr$ and $\vrho$.



\subsubsection{Cell heterogeneity}\label{sec:multi-centroid}
While the cell annotations such as cell types often correspond to distinct sub-populations of cells, significant variations may naturally exist within each sub-population. 
This means a single vector $\mX^{\uR}_{c,:}$ may not properly represent the distribution of cells within sub-population $c$. Consequently, transferring $c$ solely based on the centroid of cells that belong to $c$ may not capture these variations. To capture this intrinsic heterogeneity, we cluster each sub-population into predefined $\kappa$ smaller groups using an unsupervised learning method, and produce a total of $\kappa|\sC|$ centroids to replace the original $|\sC|$ centroids. With this definition of centroids, we treat all terms as before, except $d_{\text{A}}$, since prior information about sub-populations (and not their sub-clusters) are available. 

Note that this approach can be extended to singleton sub-clusters, in which case \methodname{} transfers the individual cells from the reference scRNA-seq data to the spatial data. However, transferring individual cells may be computationally expensive and prone to over-fitting, particularly when the reference data and the spatial data are not matched or when there is significant drop-out in the reference scRNA-seq data. In general, we treat the sub-clusters with very few cells as outliers and remove them to obtain a set $\sK_c$ of sub-clusters for sub-population $c\in \sC$. Once $\mY$ is obtained, $\sum_{k\in \sK_c}\emY_{k,i}$ determines the abundance of sub-population $c$ in spot $i$.

\subsubsection{Sparsity of composition}
As previously discussed, spatial data are either high-resolution (single-cell level) or low-resolution (multicell level). In the case of high-resolution spatial data, given that each spot corresponds to an individual cell (i.e., $n_i=1$), we expect that spots are pure (as opposed to mixed), in the sense that we prefer $\emY_{c,i}$ close to 0 or 1. In general, assuming that size of spot $i$ is $\bar{n}_i$ (i.e., $\bar{n}_i=\sum_{c\in \sC}Y_{c,i}$) and $\emY_{c,i}\in \{0,\bar{n}_i\}$, then $\emY_{c,i}=\bar{n}_i$ for exactly one category $c$ and is zero for all other categories. Consequently, for binary-valued $\mY$ we obtain
\begin{align*}
    \dcos\left(\mX^{\uS}_{i,:}, \sum\nolimits_{c\in \sC}\emY_{c,i}\mX^{\uR}_{c,:}\right) = \frac{1}{\bar{n}_i}\sum\nolimits_{c\in \sC}\emY_{c,i} \dcos\left(\mX^{\uS}_{i,:}, \mX^{\uR}_{c,:}\right),
\end{align*}
which is linear in $\mY$ for fixed $\bar{n}_i$. As linear objectives promote sparse (or corner point) solutions, we may control the level of sparsity of the solution by introducing a parameter $\theta\in[0,1]$ and redefining $d_{i}(\mY)$ as
\begin{align}
    d_i(\mY)= &(1-\theta)\dcos\left(\mX^{\uS}_{i,:}, \sum\nolimits_{c\in \sC}\emY_{c,i}\mX^{\uR}_{c,:}\right) + \frac{\theta}{\bar{n}_i}\sum\nolimits_{c\in \sC}\emY_{c,i} \dcos\left(\mX^{\uS}_{i,:} \mX^{\uR}_{c,:}\right). \label{eq:distance-location-sparse}
\end{align}
Note that a higher value for $\theta$ yields a sparser solution. Indeed, with $\theta=1$ and zero weights assigned to other objectives, the optimal solution will be completely binary. Note that $\bar{n}_i$ acts as a penalty weight and can be set to a fixed value (e.g., $n_i$).

\subsection{A fast Frank-Wolfe implementation}

We propose a solution to the \methodname{} model based on the Frank-Wolfe (FW) algorithm \cite{frank1956algorithm, jaggi2013revisiting}, which is a first-order method for solving non-linear optimization problems of the form 
$\min_{\vx\in \sX}f(\vx)$,
where $f : \mathbb{R}^n \rightarrow \mathbb{R}$ is a (potentially non-convex) continuously differentiable function over the convex and compact set $\sX$.
FW operates by replacing the non-linear objective function $f$ with its linear approximation $\tilde{f}(\vx)=f(\vx^{(0)})+\nabla_\vx f(\vx^{(0)})^{\top}(\vx-\vx^{(0)})$ at a trial point $\vx^{(0)}\in \sX$, and solving a simpler problem 
$\hat{\vx}=\arg\min_{\vx\in \sX}\tilde{f}(\vx)$
to produce an ``atom'' solution $\hat{\vx}$. The algorithm then iterates by taking a convex combination of $\vx^{(0)}$ and $\hat{\vx}$ to produce the next trial point $\vx^{(1)}$, which remains feasible thanks to convexity of $\sX$. The FW algorithm is described in Algorithm \ref{alg:fw}, in which $f(\mY)$ is the objective function in Eq. \eqref{dot:obj}. Implementation details can be found in Appendix \ref{app:fw_implementation}.

\begin{algorithm2e}[t!]
    
    Set $t=0$; find an initial solution $\mY^{(0)}$ (Appendix \ref{app:fw_initial})
	
	\While{not converged}{
		Compute gradient $\mDelta^{(t)}=\nabla_\mY f(\mY^{(t)})$ (Appendix \ref{app:fw_derivatives})

        Compute the atom solution $\hat{\mY}^{(t)}$:
        
		\For{each spot $i \in \sI$}{
		Find the current best category $\hat{c}=\arg\min_{c\in \sC} \{\Delta^{(t)}_{c,i}\}$.

        Set $\hat{\emY}^{(t)}_{c,i}=0$ for $c\ne\hat c$.
  
        If $\Delta^{(t)}_{c,i} < 0$, set $\hat{\emY}^{(t)}_{\hat{c},i}= n_i$, otherwise set  $\hat{\emY}^{(t)}_{\hat{c},i}= 1$
		}
		
		
		Update $\mY^{(t+1)}= \mY^{(t)}+\frac{2}{2+t}(\hat{\mY}^{(t)}-\mY^{(t)})$
		
		$t\leftarrow t+1$
	}
	\caption{Frank-Wolfe algorithm for \methodname{}}
	\label{alg:fw}
\end{algorithm2e}

While the \methodname{} model is not separable, its linear approximation can be decomposed to $|\sI|$ independent subproblems, one for each spot $i\in \sI$. This is because, unlike conventional OT formulations, we do not require the marginal distribution of cell sub-populations (i.e., $\sum_{i\in \sI}Y_{c,i}$) to be equal to their expected distribution (i.e., $r_c$), but have penalized their deviations in the objective function using $d_{A}$ defined in Eq. \eqref{eq:distance-abundance}. The subproblem $i$ then becomes
\begin{align*}
    \min\; & \left\{\dprod{\mY_{:,i}}{\mDelta^{(t)}_{:,i}}: \mY_{:,i}\in \mathbb{R}_+^{|\sC|}, \quad 1\le \sum\nolimits_{c\in \sC}\emY_{c,i}\le n_i\right\} 
\end{align*}
which has a simple solution. Denoting the category with smallest coefficient by $\hat{c}$, if cost coefficient of $\hat{c}$ is negative then $\emY_{\hat{c},i}=n_i$, otherwise $\emY_{\hat{c},i}=1$. Consequently, $\emY_{c,i}=0$ for all other categories.
This property of Algorithm \ref{alg:fw} enables it to efficiently tackle problems with large number of spots in the spatial data.

\subsection{Experimental setup}\label{sec:setup}
\subsubsection{Parameter setting}\label{sec:setting}

In its most general form, our multi-objective formulation for \methodname{} involves the penalty weights $\lambdac$, $\lambdag$, $\lambdas$ and $\lambdaa$ in Eq. \eqref{dot:obj}, the upper bound on size of spots $n$ in Eq. \eqref{dot:cnt}, and the spatial neighborhood parameters $\bar{w}$ and $\bar{r}$ that derive the definition of spatial pairs $\sP$ in Eq. \eqref{eq:spatial-pairs}. Here, we show how all of these parameters can be inferred from the data, hence eliminating the need for the user to tune these parameters. 

We set the penalty weights in such a way that all objectives contribute equally to the objective function. More specifically, we set  $\lambdac=\frac{|\sI|}{|\sC|}$ and  $\lambdag=\frac{|\sI|}{|\sG|}$
since $\sum_{i\in \sI} d_i(\mY)$ is in the range of 0 and $|\sI|$, while $\sum_{c\in \sC}d_c(\mY)$ and $\sum_{g\in \sG}d_g(\mY)$ are upperbounded by $|\sC|$ and $|\sG|$, respectively. 
We set the upper bound on the size of spots to $n=\frac{N}{|\sI|}$ where $N$ is the total number of cells that can fit the spatial data. Clearly, $N=|\sI|$ in high-resolution SRT since each spot is at single-cell resolution, thus $n=1$.
For the low-resolution case, we employ a generalized linear regression model to estimate $N$ (see Appendix \ref{app:fw_initial}).
We also set $\lambdas=\frac{|\sI|}{n|\sP|}$ as it is not difficult to verify that $0\le \dspatial(\mY)\le n|\sP|$ when Jensen-Shannon divergence is computed in base 2 logarithm. 
Similarly, whenever prior information about the expected abundance of sub populations (i.e., $\vr$) is available, we scale $\vr$ such that $\sum_{c\in \sC}r_c \approx N$ and set $\lambdaa=\frac{|\sI|}{N}=\frac{1}{n}$. When such information is not available, we turn off this objective by setting $\lambdaa=0$.

We set the sparsity parameter $\theta=1$ for high-resolution SRT, and set $\theta=0$ for low-resolution SRT. To capture heterogeneity of sub-populations, we clustered each sub-population $c\in \sC$ into $\kappa=10$ clusters and filtered out the sub-clusters containing less than $1\%$ of the total number of cells in $c$. To compute the distance threshold $\bar{d}$, we computed the Euclidean distance of each spot to its 8 closest spots in space\footnote{We used 8 closest neighbors to mimic the number of adjacent tiles in a 2D regular grid.}, yielding $8|\sI|$ values. We then took $\bar{d}$ as the $90^{\text{th}}$ percentile of these values. Finally, we set $\bar{w}$ to the maximum of $0.6$ and the largest value that maintains $|\sP|\le |\sI|$ to ensure meaningful spatial neighborhoods and that $\dspatial$ scales linearly in the number of spots for the sake of computational efficiency. 

For \texttt{RCTD}, \texttt{SPOTlight}, \texttt{Tangram}, and \texttt{C2L} we used the default parameters suggested by the authors with the following exceptions. For \texttt{RCTD} we set the parameter \texttt{UMI\_min} to 50 to prevent the model from removing too many cells from the data. Given the large number of cell types in the mouse MOp datasets, for \texttt{SPOTlight} we reduced the number of cells per cell type to 100 to enhance the computation time. Similarly, as \texttt{Tangram} was not able to produce results in a reasonable time for the MOp instances, we randomly selected 500 cells per cell type to reduce the computation time. For \texttt{C2L}, we used 20000 epochs to balance computation performance and accuracy. For \texttt{Seurat} and \texttt{SingleR}, we followed the package documentations, with functions used with default parameters. For \texttt{RF} we used the implementation provided in the R package \texttt{ranger} \cite{wright2015ranger} with all parameters set at their default values.

\subsubsection{Performance metrics}\label{sec:metrics}
We used three metrics for comparing the performance of different models in predicting the composition of spots. 
In our high-resolution spatial data coming from the MOp region of mouse brain, we know the cell type of each single-cell spot given as $\emP_{c,i}=1$ if spot $i$ is of type $c$, and $\emP_{c,i}=0$ otherwise. We can therefore treat the cell type prediction as a multiclass classification task.

\textit{Accuracy} is the proportion of correctly classified spots (i.e., sum of the main diagonal in the confusion matrix) over all spots.
We also use \textit{Brier Score}, also known as mean squared error, to compare the accuracy of membership probabilities produced by each model:
$$\text{Brier Score} = |\sI|^{-1}\sum\nolimits_{i\in \sI}\sum\nolimits_{c\in \sC} (\emY_{c,i}-\emP_{c,i})^2,$$
where $\emY_{c,i}$ is the probability predicted by the model that spot $i$ is of cell type $c$. As Brier Score is a strictly proper scoring rule for measuring the accuracy of probabilistic predictions \cite{gneiting2007strictly}, lower Brier Score implies better-calibrated probabilities.

Besides the cell type that each spot is annotated with, we can produce a cell type probability distribution for each spot by considering the cell type of its neighboring spots, using a Gaussian smoothing kernel of the form
\begin{align*}
    \tilde{\emP}_{c,i} =& (\sum\nolimits_{j\in \sI} \emK_{i,j})^{-1}\sum\nolimits_{j\in \sI} \emK_{i,j}\emP_{c,j},
\end{align*}
where $\emK_{i,j}=\exp\left(-\|\vx_i-\vx_j\|^2/2\sigma^2\right)$ and $\sigma$ is the kernel width parameter which we set to $0.5\bar{d}$. Note that as spot $j$ becomes closer to spot $i$, its label contributes more to the probability distribution at spot $i$.
Using these probabilities, we also introduce the \textit{Spatial Jensen-Shannon} (SJS) divergence to compare the probability distributions assigned to spots (i.e., $\mY$) with the smoothed probabilities (i.e., $\tilde{\mP}$)
\begin{align*}
    \text{SJS}=\frac{1}{|\sI|}\sum\nolimits_{i\in \sI}\djs(\mY_{:,i},\tilde{\mP}_{:,i}),
\end{align*}
where $\djs(\mY_{:,i},\tilde{\mP}_{:,i})$ is the Jensen-Shannon divergence between probability distributions $\mY_{:,i}$ and $\tilde{\mP}_{:,i}$ with base 2 logarithm as defined in Eq. \eqref{eq:jensen-shannon}.

Unlike the high-resolution spatial data, the ground truth $\emP_{c,i}$ in the low-resolution spatial data corresponds to relative abundance of cell type $c$ in spot $i$. We can therefore assess the performance of each model by comparing the probability distributions $\mP_{:,i}$ and the estimated probabilities (i.e., $\mY_{:,i}$) using Brier Score or Jensen-Shannon metrics.

\subsubsection{Data preparation}\label{sec:data-preparing}

For experiments on transferring cell types to high-resolution spatial data (Section~\ref{sec:experiment-celltype-prediction}), with each sample of the MERFISH MOp (see Appendix~\ref{sec:data_mop}), we created a reference single-cell data using all the 280,186 cells, except the cells contained in the sample, and the 254 genes to estimate the centroids of the 99 reference cell types. We further created 15 high-resolution spatial datasets for each sample (i.e., a total of 1125 spatial datasets) as follows.
To simulate the effect of number of shared features between the spatial and scRNA-seq data, we assumed that only a subset of the 254 genes are available in the spatial data by selecting the first $|\sG|$ genes, where $|\sG|\in \{50,75,100,125,150\}$ (i.e., 20\%, 30\%, 40\%, 50\%, 60\% of genes). Moreover, to simulate the effect of differences in measurement sensitivities of different technologies, we introduced random noise in the spatial data by multiplying the expression of gene $g$ in spot $i$ by $1+\beta_{i,g}$, where $\beta_{i,g}\sim U(-\varphi,\varphi)$ with $\varphi\in \{0,0.25,0.5\}$.

We produced ground truth for low-resolution MOp using the common subclass annotations between MERFISH MOp and scRNA-seq MOp \cite{yao2021transcriptomic} (see Appendix~\ref{sec:data_mop}) as follows. For each of the 75 MERFISH MOp samples, we randomly assigned each cell in the MERFISH MOp data to a cell in the scRNA-seq MOp data of the same subclass. Next, we lowered the resolution of spatial data by splitting each sample into regular grids of length 100\textmu m and aggregated the expression profiles of cells within each tile as the expression profile of the respective spots.

For experiments on estimating the expression of unmeasured genes in low-coverage spatial data (Section~\ref{sec:experiment-expression-estimation}), we matched the common capture areas of high- and low-resolution spatial data using the Hematoxylin-Eosin (H\&E) images accompanying these spatial data (Supplementary Fig. \ref{fig:xen_vis_overlap}), which corresponded to 134,664 cells in the high-resolution and 3,928 spots in the low-resolution spatial data. 
Given that the task at hand is to estimate the expression of missing genes in the high-resolution spatial data, we performed community detection on the graph of shared nearest neighbors of cells in scRNA-seq using the Leiden implementation in \cite{stuart2019comprehensive}, 
which is common practice in single-cell analysis and is used as a first step towards cell sub-population identification (note that the reference scRNA-seq does not contain cell type annotations).
This resulted in 218 clusters; we then transferred the centroids of these clusters to the high-resolution spatial data. (We also tried as high as 1000 fine-grained clusters but got essentially the same results.) 

\section{Data availability}

Publicly available single-cell RNA-seq and spatial data can be accessed via the following accession numbers or the links provided.
MERFISH data of mouse MOp \cite{zhang2021spatially} can be accessed at the Brain Image Library: \url{https://doi.org/10.35077/g.21}. 
Single-cell RNA-seq data of mouse MOp \cite{yao2021transcriptomic} and SSp \cite{yao2021taxonomy} can be accessed at the NeMO Archive for the BRAIN Initiative Cell Census Network via \url{https://assets.nemoarchive.org/dat-ch1nqb7} and \url{https://assets.nemoarchive.org/dat-jb2f34y}, respectively.
osmFISH data of mouse SSp is available at \url{http://linnarssonlab.org/osmFISH/}.
ISS and scRNA-seq data of the developing human heart \cite{asp2019spatiotemporal} is available at the European Genome-phenome Archive via accession number \href{https://ega-archive.org/studies/EGAS00001003996}{EGAS00001003996}.
Xenium, Visium and scRNA-seq data of human breast cancer \cite{janesick2022high} can be accessed at \url{https://www.10xgenomics.com/products/xenium-in-situ/preview-dataset-human-breast}.
More detailed description of these datasets can be found in
Appendix \ref{sec:data}.

\section{Code availability}
The code is open source and freely available at \url{https://github.com/saezlab/dot}.

\section{Acknowledgements}
We thank Ricardo O. Ramirez-Flores (Heidelberg University) and Zeinab Mokhtari (GSK) for their valuable discussions.

\section{Conflict of interests}
AR is supported by funding from GSK. JSR reports funding from GSK, Pfizer and Sanofi and fees/honoraria from Travere Therapeutics, Stadapharm, Astex, Pfizer and Grunenthal.

\section{Ethic statement}
The human biological samples were sourced ethically and their research use was in accord with the terms of the informed consents under an IRB/EC approved protocol.

All animal studies were ethically reviewed and carried out in accordance with European Directive 2010/63/EEC and the GSK Policy on the Care, Welfare and Treatment of Animals.

\bibliography{dot_mi}

\clearpage

\begin{appendices}

\renewcommand{\thefigure}{A\arabic{figure}}
\setcounter{figure}{0}
\setcounter{page}{1}

\section{Implementation details of the FW algorithm}\label{app:fw_implementation}


\subsection{Convergence}
Under suitable conditions, FW converges to an optimal solution in linear rate  when optimizing a convex function over a polytope domain \cite{jaggi2015global}. Given the non-convex objective function in \eqref{dot:obj}, Algorithm \ref{alg:fw} instead obtains a first-order stationary point at a rate of $O(1/\sqrt{t})$ \cite{bertsekas2016nonlinear, wai2017decentralized}. We numerically assess the convergence of Algorithm \ref{alg:fw} at iteration $t$ using the so-called ``FW-gap''
\cite{jaggi2013revisiting} 
$$\delta^{(t)} \coloneqq \sum\nolimits_{i\in \sI}\sum\nolimits_{c\in \sC} (\emY^{(t)}_{c,i}-\hat{\emY}^{(t)}_{c,i})\Delta^{(t)}_{c,i}.$$
We also implemented acceleration techniques such as averaging gradients \cite{zhang2021accelerating}, away steps \cite{jaggi2015global,garber2016linear}, and entropic regularization but did not observe substantial gains compared to our current implementation of FW. 

\subsection{Initial solution}\label{app:fw_initial}
A good quality initial solution can enhance convergence of FW. Given the multi-objective nature of our model, we produce an initial solution as convex combination of three solutions. In the first solution, for each spot $i$ we first find cell type $\hat{c}=\argmin_{c\in \sC}\{\dcos \left(\mX^{\uS}_{i,:}, \mX^{\uR}_{c,:}\right)\}$ and set $\emY_{c,i}=n_i$ if $c=\hat{c}$ and $\emY_{c,i}=0$ otherwise. Note that this solution is optimal for the sparse case when $d_i$ is the only objective. 

We derive the second solution with the goal of optimizing $d_g$ as the sole objective function. Assuming that both $\mX^\uS$ and $\mX^\uR$ are count matrices, we can approximate minimizing $d_g$ by solving a non-negative least squares
\begin{align*}
    \min_{\mY \ge \vzero}\; \|\mY^{\top}\mX^\uR-\mX^\uS\|_2^2.
\end{align*}
To derive a fast solution, we note that all entries of $\mX^\uS$ and $\mX^\uR$ are non-negative. Therefore, a generalized linear regression with the non-negativity constraints relaxed yields a solution $\mY$ in which $\emY_{c,i} > 0$ for at least one $c$ for each $i$. Finally, adding a ridge penalty to account for the cases when $\mX^\uR$ is not full-rank (which typically happens when number of genes is less than number of sub-populations), we obtain the solution
\begin{align}
    \mY = \left(\mX^\uR {\mX^\uR}^{\top} + \mI_{|C|}\right)^{-1} {\mX^\uR}{\mX^\uS}^{\top}, \label{eq:least-squares-solution}
\end{align}
and set the negative entries of $\mY$ to $0$. Given that $|\sC|$ is typically small, the matrix inversion in Eq. \eqref{eq:least-squares-solution} can be done easily. Moreover, given that $\mX^\uS$ and $\mX^\uR$ are count matrices, $\sum_{c}\sum_{i}Y_{c,i}$ gives an estimate on the total number of cells that can fit in S.

In the third solution, we simply set $\emY_{c,i}=\frac{r_c}{\sum_{c'}r_{c'}}n$ for each $i$ and $c$. Note that this solution is optimal for $d_{\text{A}}$. We then set the initial solution as the convex combination of these three solutions with weights 0.4, 0.4, 0.2, respectively.

\subsection{Derivatives}\label{app:fw_derivatives}
To find the derivatives of $d_i(\mY)$ and $d_c(\mY)$, defined in Eq. \eqref{eq:distance-location} and Eq. \eqref{eq:distance-centroid}, we introduce auxiliary quantities $\bar{\mX}^{\uS}\coloneqq \mY^{\top} \mX^{\uR}$ and $\bar{\mX}^{\uR}\coloneqq \mY \mX^{\uS}$ to denote the expressions transferred through $\mY$ to spots and cell sub-populations, respectively. 
Derivatives for $d_i(\mY)$ and $d_c(\mY)$ can then be calculated as:
\begin{align*}
    \frac{\partial d_i}{\partial \emY_{c,i}}=\frac{1}{\|\mX^{\uS}_{i,:}\|}\dprod{\mX^{\uR}_{c,:}}{\mT^{\uS}_{i,:}},\qquad
    \frac{\partial d_c}{\partial \emY_{c,i}}= \frac{1}{\|\mX^{\uR}_{c,:}\|}\dprod{\mX^{\uS}_{i,:}}{\mT^{\uR}_{c,:}},
\end{align*}
where
\begin{align*}
    \emT^{\uS}_{i,g}=&\frac{-1}{2d_i(\mY)}\left(\frac{\emX^{\uS}_{i,g}}{\|\bar{\mX}^{\uS}_{i,:}\|}-\frac{\bar{\emX}^{\uS}_{i,g}}{\|\bar{\mX}^{\uS}_{i,:}\|^3}\dprod{\mX^{\uS}_{i,:}}{\bar{\mX}^{\uS}_{i,:}}\right),\\
    \emT^{\uR}_{c,g}=&\frac{-1}{2d_c(\mY)}\left(\frac{\emX^{\uR}_{c,g}}{\|\bar{\mX}^{\uR}_{c,:}\|}-\frac{\bar{\emX}^{\uR}_{c,g}}{\|\bar{\mX}^{\uR}_{c,:}\|^3}\dprod{\mX^{\uR}_{c,:}}{\bar{\mX}^{\uR}_{c,:}}\right).
\end{align*}
Similarly, we may derive the derivatives for $d_g(\mY)$ defined in Eq. \eqref{eq:distance-gene} via
\begin{align*}
    \frac{\partial d_g}{\partial \emY_{c,i}}=&
    \frac{-1}{2d_{g}(\mY)}\frac{\emX^{\uR}_{c,g}}{\|\mX^{\uS}_{:,g}\|}\left(\frac{\emX^{\uS}_{i,g}}{\|\bar{\mX}^{\uS}_{:,g}\|}-\frac{\emY_{c,i}}{\|\bar{\mX}^{\uS}_{:,g}\|^3}\dprod{\mX^{\uS}_{:,g}}{\bar{\mX}^{\uS}_{:,g}}\right)
\end{align*}
The derivatives for $\dspatial$ defined in Eq. \eqref{eq:distance-spatial} can be computed as
\begin{align*}
    \frac{\partial \dspatial}{\partial \emY_{c,i}}= \frac{1}{2}\sum_{j\in \sI: (i,j)\in \sP \text{ or } (j,i)\in \sP} \log\left(\frac{2\emY_{c,i}}{\emY_{c,i}+\emY_{c,j}}\right).
\end{align*}
Finally, the derivatives for $d_{\text{A}}$ defined in Eq. \eqref{eq:distance-abundance} can be calculated as:
\begin{align*}
    \frac{\partial d_{\text{A}}}{\partial \emY_{c,i}}=\frac{1}{2}\log\left(\frac{2\rho_c}{\rho_c+r_c}\right).
\end{align*}

\section{Connection to Fused Gromov-Wasserstein Optimal Transport}\label{app:connection-to-ot}

As discussed in the main body of the paper, our formulation can be viewed as a generalization of Optimal Transport. Here, we elaborate on connections between our formulation and standard OT formulations and highlight the distinct features of our model that separate our formulation from them. An OT formulation in its most basic form for assigning cell sub-populations to spatial locations can be  expressed as the following optimization problem:
\begin{align}
    \min_{\mZ \ge \vzero} \quad & \sum_{c\in \sC}\sum_{i\in \sI} \emC_{c,i}\emZ_{c,i} \label{eq:ot-obj}\\
    \st \quad & \sum_{c\in \sC} \emZ_{c,i} = p_i \qquad \forall i\in \sI \label{eq:ot-i}\\
    & \sum_{i\in \sI} \emZ_{c,i} = q_c \qquad \forall c\in \sC, \label{eq:ot-c}
\end{align}
where $\vp$ and $\vq$ are given marginal distributions for cell sub-populations and spots, respectively, and $\mC$ is the transportation cost matrix which can be computed as the dissimilarity between expression profile of sub-populations in R and spots in S. 

We first note that the linear cost function in Eq. \eqref{eq:ot-obj} is akin to our location-wise cost function $d_i$ in the sparse case when $\emC_{c,i}=\dcos(\mX^\uR_{c,:}, \mX^\uS_{i,:})$. More precisely, $d_i(\mZ) = \sum_{c\in \sC} \emC_{c,i}\emZ_{c,i}$ when the sparsity parameter $\theta$ in Eq. \eqref{eq:distance-location-sparse} is set to 1. However, there are major differences between $d_i$ and the linear cost function which make our distance function $d_i$ more suitable for the task at hand:
\begin{enumerate}[(i)]
    \item First, note that $\emC_{c,i}$ is computed by assuming that all of location $i$ is occupied by a single sub-population $c$. Therefore, a linear cost function cannot capture the low resolution case as spots in the low-resolution SRT are comprised of multiple cells that potentially belong to different sub-populations.
    \item The second difference between $d_i$ and the linear cost function is that $d_i$ is indifferent to the size of spots in the low-resolution case thanks to the scale invariance property of our $\dcos$ distance function. In contrast, the linear cost function pushes the size of all spots to the lower limit. More precisely, if we relax \eqref{eq:ot-c} and replace \eqref{eq:ot-i} with a two-sided bounded constraint $1\le \sum_{c\in \sC} \emZ_{c,i} \le n_i$, then $\sum_{c\in \sC} \emZ_{c,i}=1$ at any optimal solution. This means a standard OT formulation (even a partially unbalanced Fused Gromov-Wasserstein formulation; see below) cannot distinguish between the size of different spots.
    \item Finally, when reliable information about the abundance of sub-populations is not available, even a partially unbalanced OT formulation may not be appropriate and the OT formulation results in a trivial solution in which each spot gets assigned to its closest sub-population independently of other spots. Note that our centroid distance function $d_c$ and gene map distance function $d_g$ defined in Eq. \eqref{eq:distance-centroid} and Eq. \eqref{eq:distance-gene}, respectively, prevent such a trivial solution even when no prior information about the abundance of sub-populations is available.
\end{enumerate}

The second link between our formulation and variants of OT can be characterized via the Fused Gromov-Wasserstein (FGW) formulation, a variant of OT for matching structured data. In our application, given $\mM^\uR$ and $\mM^\uS$ as metrics in the space of R and S, which denote the pairwise dissimilarity between elements of R and S, respectively, FGW combines the linear cost $\sum_{c\in \sC}\sum_{i\in \sI} \emC_{c,i}\emZ_{c,i}$ with the \textit{2-Gromov-Wasserstein} distance \cite{memoli2011gromov} and replaces the objective function in Eq. \eqref{eq:ot-obj} with
\begin{align}
    \alpha \sum_{c\in \sC}\sum_{i\in \sI} \emC_{c,i}^2\emZ_{c,i} + (1-\alpha) \sum_{c\in \sC}\sum_{k\in \sC}\sum_{i\in \sI}\sum_{j\in \sI} \emZ_{c,i}\emZ_{k,j}\left(M^\uR_{c,k}-M^\uS_{i,j}\right)^2 \label{eq:fgw-obj}
\end{align}
for some $\alpha\in[0,1]$. From this perspective, the GW distance component of \eqref{eq:fgw-obj} can capture the spatial relations between spots. In the following, we show how our spatial distance function $\dspatial$ defined in Eq. \eqref{eq:distance-spatial} is related to this distance function for a particular choice of metrics $\mM^\uR$ and $\mM^\uS$. 

\begin{proposition}\label{prop:gw-simplification} 
Let $\beta=\sum_{i\in \sI}\sum_{j\in \sI}(1-\emM^{\uS}_{i,j})^2 p_i p_j$. Assuming that $\mM^\uR$ is a discrete metric so that $\emM^{\uR}_{c,c}= 0$ and $\emM^{\uR}_{c,k}= 1$, for $c,k\in \sC$, $c\ne k$, then
\begin{align*}
    \text{GW}(\mZ) = \beta+\sum\nolimits_{i\in \sI}\sum\nolimits_{j\in \sI}\left(2M^{\uS}_{i,j}-1\right)\dprod{\mZ_{:,i}}{\mZ_{:,j}}
\end{align*}
\end{proposition}
\begin{proof}
Given $\emM^{\uR}_{c,k}=1$ for $c\ne k$ and $\emM^{\uR}_{c,c}=0$, we obtain
\begin{align*}
    \text{GW}(\mZ)=&\sum\limits_{i\in \sI}\sum\limits_{j\in \sI}\sum\limits_{c\in \sC}\left(\emM^{\uS}_{i,j}\right)^2\emZ_{c,i}\emZ_{c,j}+\sum\limits_{i\in \sI}\sum\limits_{j\in \sI}\sum\limits_{c\in \sC}\sum\limits_{k\in \sC, k\ne c} \left(1-\emM^{\uS}_{i,j}\right)^2\emZ_{c,i}\emZ_{k,j}\\
    =&\sum\limits_{i\in \sI}\sum\limits_{j\in \sI}\sum\limits_{c\in \sC}\left(\left(\emM^{\uS}_{i,j}\right)^2-\left(1-\emM^{\uS}_{i,j}\right)^2\right)\emZ_{c,i}\emZ_{c,j}+\sum\limits_{i\in \sI}\sum\limits_{j\in \sI}\sum\limits_{c\in \sC}\sum\limits_{k\in \sC} \left(1-\emM^{\uS}_{i,j}\right)^2\emZ_{c,i}\emZ_{k,j}\\
    =& \sum\limits_{i\in \sI}\sum\limits_{j\in \sI}\left(2\emM^{\uS}_{i,j}-1\right)\dprod{\mZ_{:,i}}{\mZ_{:,j}} + \beta,
\end{align*}
where we have used $\beta=\sum\limits_{i\in \sI}\sum\limits_{j\in \sI}\left(1-\emM^{\uS}_{i,j}\right)^2\sum\limits_{c\in \sC}\sum\limits_{k\in \sC} \emZ_{c,i}\emZ_{k,j}=\sum\limits_{i\in \sI}\sum\limits_{j\in \sI}\left(1-\emM^{\uS}_{i,j}\right)^2 p_i p_j$
since $\sum\limits_{c\in \sC} \emZ_{c,i}=p_i$ and $\sum\limits_{k\in \sC} \emZ_{k,j}=p_j$.
\end{proof}

Observe that $\dprod{\mZ_{:,i}}{\mZ_{:,j}}$ measures similarity between composition of spots $i$ and $j$.
Consequently, for a discrete metric $\mM^\uR$ (i.e., when sub-populations are radically different), minimizing $\text{GW}(\mZ)$ encourages spots $i$ and $j$ to acquire similar compositions when $2M^{\uS}_{i,j}-1 > 0$, discourages spots $i$ and $j$ from acquiring similar compositions when $2M^{\uS}_{i,j}-1 < 0$, and is indifferent to the composition of spots $i$ and $j$ when $2M^{\uS}_{i,j}-1 = 0$.

To produce a metric $\mM^{\uS}$ that captures the dissimilarity of spots in terms of their locations and expressions, we define $\mD^1_{i,j}$ and $\mD^2_{i,j}$ to represent distance of spots $(i,j)$ with respect to their locations and expressions, respectively
\begin{align*}
    \emD^1_{i,j} &= \1_\mathrm{condition}\left(\|\vx_i-\vx_j\|> \bar{d}\right)\\
    \emD^2_{i,j} &=\dcos\left(\mX^{\uS}_{i,:},\mX^{\uS}_{j,:}\right),
\end{align*}
where $\bar{d}$ is a given distance threshold, and $\emD^2_{i,j}$ is computed with respect to all genes in S (i.e., $\sG^{\uS}$). Finally, we take $\mM^{\uS}$ to be the average of $\mD^1$ and $\mD^2$:
\begin{align}
    \mM^{\uS} = (\mD^1+\mD^2)/2\label{eq:distance-matrix-S}
\end{align}

\begin{remark}
$\mM^{\uS}$ is a metric in the domain of S, since both $\mD^1$ and $\mD^2$ are metrics.
\end{remark}

\begin{remark}
With the definition of $\mM^{\uS}$ in Eq. \eqref{eq:distance-matrix-S} and $\mM^{\uR}$ a discrete metric,
$\text{GW}(\mZ)$ encourages adjacent spots to attain similar compositions if their expressions are similar, (ii) discourages distant spots from attaining similar compositions if their expressions are different, and (iii) is indifferent to pair $(i,j)$ when $i$ and $j$ are distant or different in expressions, but not both.
\end{remark}

From this perspective, our spatial distance function $\dspatial$ defined in Eq. \eqref{eq:distance-spatial} specializes $\text{GW}(\mZ)$ to encouraging adjacent spots to attain similar compositions if their expressions are similar. Note that our definition of set of spatial pairs $\sP$ given in Eq. \eqref{eq:spatial-pairs} uses the same distance threshold $\bar{d}$. However, given the non-convex and quadratic nature of $\text{GW}(\mZ)$, our $\dspatial$ distance function is computationally more appealing as it is convex and scales linearly with the number of spots. 

\section{Datasets}\label{sec:data}
\subsection{Mouse Primary Motor Cortex (MOp)}\label{sec:data_mop}

We used the spatially resolved cell atlas of the MOp recently generated using multiplexed error-robust fluorescence in situ hybridization (MERFISH) technology and made publicly available
by \cite{zhang2021spatially}. The processed dataset contains normalized RNA counts of 254 genes and coordinates of the boundaries of a total of 280,186 segmented cells across 75 samples in the MOp of two adult mice, with the number of cells within each sample ranging from 1000 to 7500 cells.
We computed the $(x,y)$ coordinates of the center of each cell by taking the average of the coordinates of its boundary. The study also identifies 99 trasncriptionally distinct cell types by community detection applied on a cell similarity graph. The clustering resulted in 39 excitatory neuronal cell types (clusters), 42 inhibitory neuronal cell types, 14 non-neuronal cell types, and four other cell types.

The corresponding scRNA-seq data comes from a cell atlas of the MOp \cite{yao2021transcriptomic}.
We used the scRNA-seq dataset \texttt{scRNA\_10X\_v2\_A},
which contains 145,748 cells and 100 cell types. After removing the unannotated cells and low quality cell types (as categorized in the study), we retrieved 124,330 cells and 90 distinct cell types. For computational efficiency, we also selected the top 5,000 variable genes according to their means and variances \cite{stuart2019comprehensive}. 

\subsection{Mouse Primary Somatosensory Cortex (SSp)}\label{sec:data_ssp}
Similar to MOp, another well-studied tissue area is the primary somatosensory cortex area (SSp). Here, we used high-resolution spatial data coming from the osmFISH platform \cite{codeluppi2018spatial}, which contains measurements of 33 genes across 4,837 cells, as well as annotations based on 11 major cell types.
For reference scRNA-seq data with matched cell types, we used the annotations independently generated by \cite{yao2021taxonomy} using 5,392 single cells in the same SSp region. 

\subsection{Developing Human Heart}\label{sec:data_heart}

For the developing human heart, we used subcellular spatial data generated by the ISS platform 
\cite{asp2019spatiotemporal}, which contains tissue sections from human embryonic cardiac samples collected at different times. We selected the \texttt{PCW6.5} slide which contains measurements of 69 genes across 17,454 cells as well as annotations of 12 major cell types. The same study also provides scRNA-seq data for a similar slide, which contains matched cell types for 3,253 cells.

\subsection{Human Breast Cancer}\label{sec:data_breast_cancer}
Breast cancer is a complex disease with significant cellular and molecular heterogeneity. 
We used the spatial data from breast cancer tumor microenvironment produced by the 10X Xenium In Situ technology \cite{janesick2022high}. The dataset is unique in that it contains both high-resolution (Xenium) and low-resolution (Visium) spatial data of serial sections from the same tissue. The high-resolution data contains two replications produced by the recent 10X Xenium In Situ technology. We used  \texttt{Xenium\_FFPE\_Human\_Breast\_Cancer\_Rep1}, which contains the spatial information of 313 genes for 167,782 cells. The low-resolution spatial dataset is produced by the  10X Visium Spatial Transcriptomics technology, which contains the spatial information of 18,000 genes for 4,992 multicell spots. 
The dataset also contains the dissociated scRNA-seq data coming from a tissue section adjacent to the tissue sections used for Visium and Xenium workflows. We used the \texttt{Single Cell Gene Expression Flex (FRP)} data which contains expression of 18,000 genes across 30,365 cells.

\begin{figure}[h!]
    \centering
    \includegraphics[width=0.5\textwidth]{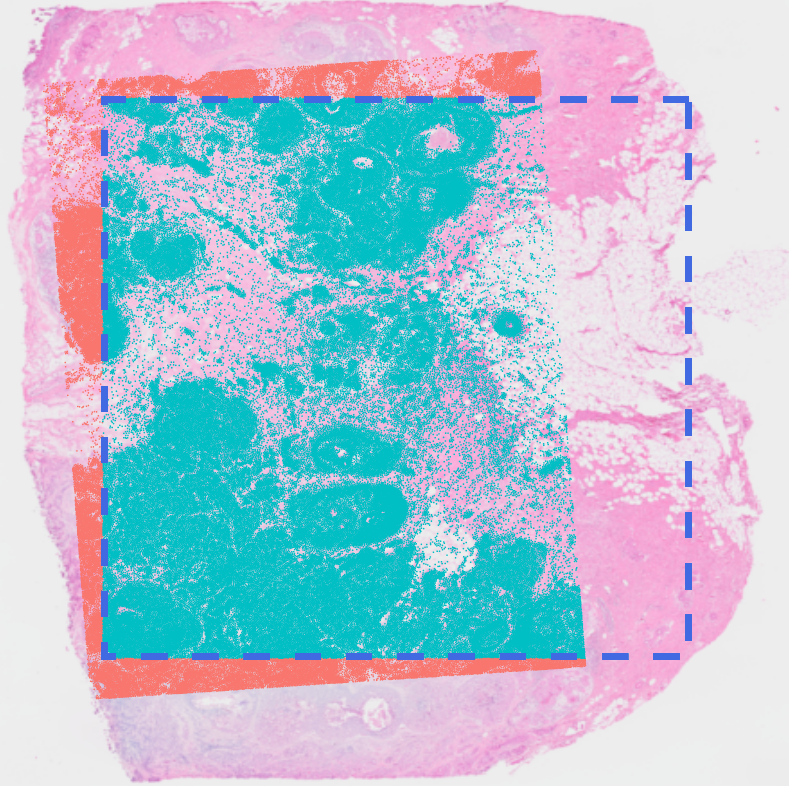}
    \caption{Common region (cyan) in the capture areas of Visium (dashed blue lines) and Xenium (dark orange) in human breast cancer. The pink region is the H\&E image accompanying Visium.}
    \label{fig:xen_vis_overlap}
\end{figure}

Fig.~\ref{fig:xen_vis_overlap} illustrates the common capture areas of Visium and Xenium tissues.

\end{appendices}

\end{document}